\documentclass[referee]{raa}
\usepackage{graphicx,times}
\usepackage{natbib}
\usepackage{multirow}
\usepackage{amssymb,amsmath}
\usepackage{enumitem}

\bibpunct{(}{)}{;}{a}{}{,}

\usepackage[a4paper=true,dvipdfm=true,pagebackref=true]{hyperref}
\hypersetup{pdftitle = The title of my PDF, pdfauthor = My name, pdfsubject= The subject, pdfkeywords = keyword1 keyword2 keyword3}
\hypersetup{colorlinks = true, linkcolor = green, anchorcolor = red, citecolor = blue, filecolor = red, pagecolor = red, urlcolor = red}

\begin{document}

   \title{Strange quark star in dilaton gravity}

 \volnopage{ {\bf 20XX} Vol.\ {\bf X} No. {\bf XX}, 000--000}
   \setcounter{page}{1}

   \author{A.R. Peivand\inst{1}, K. Naficy\inst{1}, G.H. Bordbar\inst{2,3,4}
\footnote{E-mail:
ghbordbar@shirazu.ac.ir}}

   \institute{ Department of Physics, University of Birjand, PO Box 615/97175, Birjand, South Khorasan, Iran; {\it naficy@birjand.ac.ir}\\
        \and
             Department of Physics, Shiraz University, Shiraz 71454, Iran\footnote{Permanent address}\\
	\and
Research Institute for Astronomy and Astrophysics of Maragha, P.O. Box 55134-441, Maragha, Iran\\
\and
Department of Physics and Astronomy, University of Waterloo, 200 University Avenue West, Waterloo, Ontario, N2L3G1, Canada\\
\vs \no
   {\small Received 20XX Month Day; accepted 20XX Month Day}
}

\abstract{In this work, we have first obtained the hydrostatic equilibrium equation in dilaton gravity. Then we have examined some of the structural characteristics of strange quark star in dilaton gravity with a background of Einstein gravity. We have shown that the variations of dilaton parameter do not affect the maximum mass of quark star, while the variations of the cosmological constant lead to change in the structural characteristics of the quark star. We have investigated the stability of strange quark stars that studied by MIT Bag model, in dilaton gravity. We have also provided limiting values for the dilaton field parameter and the cosmological constant.We have also studied the effects of dilaton gravity on the other properties of quark star such as the mean density and gravitational redshift.
We have concluded that the last reported value for cosmological constant does not affect on maximum mass of strange quark star.
}

   \authorrunning{A.R.Peivand et al. }            
   \titlerunning{Strange quark star in dilaton gravity}  
   \maketitle

%
\section{Introduction}
\label{sec:intro}

Einstein's general relativity (GR) theory well explains phenomena and events within the solar system. Tolman, Oppenheimer and Volkoff (TOV) ~(\citealt{ tolman1939static, oppenheimer1939massive}) obtained the first hydrostatic equilibrium equation(HEE) from the solution of the Einstein field equation. So far, the structure of compact objects, such as neutron stars and quark stars, that have general relativistic properties due to their great density has been obtained by many authors~(\citealt{silbar2005erratum, bordbar2006structure,bordbar2016neutron,narain2006compact}) through the numerical solution of the TOV equation.\\
In recent years, studies outside the solar system have led to the emergence of new theories and observations such as accelerated expansion of the universe~(\citealt{knop2003new,perlmutter1999measurements,riess1998observational,tonry2003cosmological}), which GR is incapable of explain, and also the absence of a comprehensive gravitational theory in quantum scales has led to a lot of attention to the emergence of new gravitational theories. Theories that include GR and can explain the new cosmological phenomena and discoveries that GR cannot explain and interpret well.

Adding new terms, including higher-order curvature invariants and scalar fields, to Einstein Lagrangian yields theories known as modified gravity. One of these new modified gravities is to consider the dilaton scalar field and its potential in the universe. The scalar field was used to justify the inflation phenomenon and also to describe cold dark matter~(\citealt{cho1990unified}), which is a type of dark matter. Recently, the study of the structure of compact objects in modified gravity such as neutron stars~(\citealt{astashenok2016neutron,hendi2015dilatonic}), black holes~(\citealt{chan1995charged,hendi2016charged}) and quark stars~(\citealt{astashenok2016neutron,astashenok2015nonperturbative}) has been widely considered.

 In order to obtain the structural characteristics of compact objects in the new gravities, we use the action related to the proposed gravitational theory. By varying this action and taking the principle of least action into account, we obtain the field equation for the considered gravity. By solving this equation and using the conservation of the momentum-energy tensor, we obtain HEE of the new theory. Then, by using the equation of state (EoS) of the matter of the compact object, we solve the obtained HEE equation numerically and obtain the maximum mass and radius of the considered star. The HEE in modified gravity is considered by several authors: HEE in Gauss-Bonnet gravity~(\citealt{momeni2015tolman}), HEE in $f(G)$ gravity ~(\citealt{sharif2016static}), HEE in~$f(R)$~gravity~(\citealt{astashenok2016neutron,astashenok2013further,arapouglu2011constraints}), HEE in gravity's rainbow~~(\citealt{hendi2016modified}), HEE in massive gravity~(\citealt{katsuragawa2016relativistic,hendi2017neutron})

After Witten conjectured hypothesis of strange quark matter  (\citealt{witten1984witten,farhi1984strange}), it has been shown that a new class of compact object may exist that have been composed from strange quark matter, known as strange quark stars(SQS) or strange stars(SS) (\citealt{haensel1986strange,alcock1986strange,alcock1986model,alcock1988exotic,glendenning1990nk}). Strange quark stars composed from up($u$), down($d$) and strange($s$) quarks along with a small number of electrons that are in $\beta$-equilibrium. The best candidates for being SQS are some of the observed compact objects that could not be compatible with neutron star model, such as X-ray pulsar {LMC X-4} with $M=1.29\pm0.05\,M_{\odot}$ ~(\citealt{gangopadhyay2013strange,paul2011relativistic,rawls2011refined}), X-ray burster {4U 1608-52} ~(\citealt{paul2011relativistic,bombaci1997observational,dey1998strange}) with $M=1.74\pm0.14\,M_{\odot}$ ~(\citealt{guver2010mass}) and millisecond pulsar {J1614-2230} ~(\citealt{paul2011relativistic,deb2017relativistic}) with $M=1.97\pm0.04\,M_{\odot}$ ~(\citealt{demorest2010two}). In this work, the considered quark star is made of strange quark matter from the center up close to the surface. Here, we want to calculate some bulk properties of SQS by using modified hydrostatic equilibrium equation in dilaton gravity and considering MIT Bag model to obtain equation of state of SQM. This paper is divided into six sections.
In Sec.~\ref{sec:HEE}, we propose a new procedure to find HEE in dilaton gravity on the background of Einstein gravity. In Sec.~\ref{sec:Correct}, we present the HEE in dilaton gravity as a correction of the HEE in Einstein gravity(correction of TOV).
In Sec.~\ref{sec:structure}, we obtain the maximum mass of the quark star from the numerical solution of the HEE obtained in Secs.~\ref{sec:HEE} and \ref{sec:Correct} and compare our results with the results obtained in GR. In Sec.~\ref{sec:dyst}, we have investigate causality condition and dynamical stability of SQS in dilaton gravity. Our conclusions are drawn in the~\ref{sec:con}th section.

\section{HEE in dilaton gravity}
\label{sec:HEE}
The total action of dilaton gravity($S_{total}$) is as follows
\begin{equation}
S_{total}=\frac{1}{2\kappa}\int d^{4}x\sqrt{-g}\left(R-2g^{\mu\nu}\partial_{\mu}\Phi\partial_{\nu}\Phi-V(\Phi)\right)+S_{matter},
\label{eq:act}
\end{equation}
in which Einstein's gravity is given as background.
In this action, $R$, $\Phi$ and $V(\Phi)$ are Ricci scalar, dilaton field and the potential of this field, respectively. $S_{matter}$ is the action related to the matter, which is considered to be a perfect fluid. In order to find the dilaton field equation, we vary Eq.~\ref{eq:act} with respect to the dilaton field $\Phi$ and the metric tensor $g_{\mu\nu}$. Using the least action principle, field equations are obtained as follows
\begin{equation}\label{eq:dfe1}
G_{\mu\nu}=\kappa T _{\mu\nu}+\left(2\partial_{\mu}\Phi\partial_{\nu}\Phi-g_{\mu\nu}\partial_{a}\Phi\partial^{a}\Phi-\frac{1}{2}g_{\mu\nu}V(\Phi)\right)
\end{equation}
\begin{equation}\label{eq:dfe2}
\Box\Phi-\frac{1}{4}\frac{\partial V}{\partial\Phi}=0
\end{equation}
Where $G_{\mu\nu}$ and $\kappa=\frac{8\pi G}{c^4}$ are the Einstein tensor and Einstein gravitational constant, respectively. $T_{\mu\nu}$ is the energy-momentum tensor associated with the perfect fluid. We consider that the potential of dilaton field consists of two Liouville terms.
\begin{equation}\label{V}
V(\Phi)=2\Lambda_{0}e^{2\xi_{0}\Phi}-2\Lambda e^{2\xi\Phi}
\end{equation}
Potentials of this type have been previously studied to solve dilaton black hole field equations~(\citealt{chan1995charged,dehghani2005charged}).

In order to find a static solution of Eqs.~\ref{eq:dfe1} and \ref{eq:dfe2}, we assume the four-dimensional spacetime metric as
\begin{equation}\label{eq:metric}
{ds}^{2}=-B \left( r \right) {dt}^
{{{2}}}+{\frac {{dr}^{2}}{A \left( r \right) }}+R^{2}(r){r}^{2}{d\Omega}^2
\end{equation}
in which $B(r)$, $A(r)$, and $R(r)$ are unknown functions to be determined and ${d\Omega}^2=\left( {{d}}{\theta}^{{{2 }}}+\sin^{2}\theta\,{{d}}{\phi}^{2}\right)$. We consider $R(r)$ as the ansatz which has the form
\begin{equation}
R(r)=e^{\alpha\Phi(r)}
\end{equation}
This ansatz was first used to investigate answers of field equations of charged dilaton black strings~(\citealt{dehghani2005charged}). Recently, this ansatz has been used by some authors ~(\citealt{sheykhi2006asymptotically,hendi2015dilatonic}) to study properties of dilaton gravity. It is important to note that when $\alpha=0$ the ansatz becomes $R(r)=1$ and dilaton gravity turns to Einstein gravity. By making this ansatz and using Eq.~\ref{eq:dfe2} and introduced metric in Eq.~\ref{eq:metric} we have
\begin{equation}\label{eq:phiii}
\Phi(r)=\frac{\alpha}{1+\alpha^{2}}\ln\left(\frac{b}{r}\right)
\end{equation}
where b is an arbitrary constant.

For a perfect fluid, the general form of energy-momentum tensor is given by
\begin{equation}\label{eq:emt}
T^{\mu \nu }=\bigl(\rho c^{2}+P\bigr) u^{\mu }u^{\nu }+Pg^{\mu \nu },
\end{equation}
In these relations, $P$ and $\rho$ are the pressure and energy density of the perfect fluid from the viewpoint of the local observer respectively and $u^{\mu }$ is the local fluid velocity 4-vector. Using the metric Eqs.~\ref{eq:metric}and \ref{eq:emt}, the energy-momentum tensor for the perfect fluid has the form
\begin{equation}\label{eq:emtdiag}
{T^{a}}_{b}=diag(-\rho c^2,P,P,P).
\end{equation}

To solve dilaton field equation, with the metric Eq.~\ref{eq:metric} and energy-momentum tensor Eq.~\ref{eq:emtdiag}, we obtain the components of Eq.~\ref{eq:dfe1} as follows
\begin{equation}\label{eq:comp1}
\kappa\rho{c}^{2}= -\Lambda\,{\gamma}^{{\alpha}^{2}}-{\frac { A'}{r K_{{1,1}}}}-{\frac {AK_{{1,-3}}}{{r}^{2}{{{K}_{1,1}^2}}}}+{\frac {\gamma\,{\alpha}^{2}}{K_{{-1,1}}{b}
^{2}}}
+{\frac {1}{{r}^{2}{\gamma}^{{\alpha}^{2}}}}
\end{equation}
\begin{equation}\label{eq:comp2}
\kappa P =\Lambda{\gamma}^{{\alpha}^{2}}+\frac {A B'}{r B K_{{1,1}}}+{\frac {AK_{{1,-3}}}{{r}^{2}{{{K}_{1,1}^2}}}}-{\frac {\gamma\,{\alpha}^{2}}{K_{{-1,1}}{b}^{2}}}-{\frac {1
}{{r}^{2}{\gamma}^{{\alpha}^{2}}}}
\end{equation}
\begin{equation}\label{eq:comp3}
\begin{split}
\kappa  P & =\Lambda  {\gamma}^{{\alpha}^{2}}+{\frac {A  {{
 B' }}}{2rBK_{{1,1}}}}-  {\frac {2{\alpha}^{2}A}{{r}^{2}{{K}_{1,1}^2}}
}-{\frac {\gamma  {\alpha}^{2}}{K_{{-1,1}}{b}^{2}}}- {\frac {A{
  {{    B' }}}^{2}}{4{B}^{2}}}\\ &+ {\frac {  {{    A' }}
  {{    B' }}}{4B}} +{\frac {  {{    A' }}}{2r K_{{1,1}}}}+
{\frac {A  {{    B'' }}}{2B}}
\end{split}
\end{equation}
where
\begin{eqnarray}
  K_{i,j}\equiv i+j\alpha^2 , \ \ \ \ \ \gamma\equiv\left(\frac{b}{r}\right)^{\frac{2}{{K}_{1,1}}} \nonumber
\end{eqnarray}
  and the prime and double-prime are the keys that denote the first and second derivatives with respect to $r$ respectively. Furthermore, by using the conservation of the energy-momentum tensor ${T^{\mu\nu}}_{;\mu}=0$ we have
\begin{equation}\label{eq:cemt}
 \frac{dP}{dr}=-(\rho c^2+P)(\frac{B'}{B})
\end{equation}

It is notable that in the equations \ref{eq:comp1} to \ref{eq:comp3}, we set
$\Lambda$ to be a free parameter and behaves same as cosmological constant. On the other hand, $\Lambda_{0}$, $\xi_{0}$ and $\xi$ are constants that are chosen as follows to solve Eqs. \ref{eq:dfe1} and \ref{eq:dfe2},
\begin{eqnarray}
\Lambda_{0}=\frac{\alpha^2}{b^2 K_{-1,1}},\ \ \ \ \   \xi_{0}=\frac{1}{\alpha},\ \ \ \ \    \xi=\alpha. \nonumber
\end{eqnarray}
The function $A(r)$ is obtained by integrating Eq.~\ref{eq:comp1} as the form
\begin{equation}\label{eq:ar}
\begin{split}
  A(r)=&\left({
\frac {{\gamma}^{-{\alpha}^{2}}}{{r}^{2}K_{{1,-1}}}}-{\frac {\Lambda\,{\gamma}^{{\alpha}^{2}}}{3K_{{1,-1}}}}-{\frac {\gamma\,{
\alpha}^{2}}{{{K}_{1,-1}^2}{b}^{2}}} \right) {r}^{2}{K}_{1,1}^2\\&-
\frac{\kappa\,{c}^{2}K_{{1,1}}}{ {r}^{{\frac {K_{{1,-3}}}{K_{
{1,1}}}}} }\int \!{r}^{{\frac {2K_{{1,-1}}}{K_{{1,1}}}
}}\rho \left( r \right) \,{\rm d}r
\end{split}
\end{equation}
 If we replace $\alpha=0$ in this equation we see that $A(r)$ in dilaton field reduces to known $A(r)=(1-{\frac {2\,GM}{r{c}^{2}}}-\frac{\Lambda}{3}\,{r}^{2})$ in Einstein-$\Lambda$ gravity~(\citealt{stuchlik2000spherically,balaguera2005astrophysical,bohmer2005does}),
  therefore, we can write
${4\pi\int \!{r}^{{\frac {2K_{{1,-1}}}{K_{{1,1}}}
}}\rho \left( r \right) \,{\rm d}r = M_{eff}(r , \alpha)}$ that $M_{eff}(r ,\alpha)=\int \!4\pi\rho \left( r \right){R}_{eff}^{2}
 \,{\rm d}R_{eff}$ is the effective mass with the corresponding effective radius that given by
 $R_{{{\it eff}}}=\sqrt [3]{{\frac {3K_{{1,1}}}{K_{{3,-1}}}}}\,
{r}^{\left({\frac {K_{{3,-1}}}{3K_{{1,1}}}}\right)}$.
One can obtain $\frac{B'}{B}$ by substituting Eq.~\ref{eq:ar} into Eq.~\ref{eq:comp2} ,then replace $\frac{B'}{B}$ into Eq.~\ref{eq:cemt}. With this replacement and using $\frac{dP}{dr}=\frac{dP}{dR_{eff}}\frac{dR_{eff}}{dr}$ we obtain dilaton HEE as
\begin{equation}
 \begin{split}
 \frac{{ dP}}{{dR}_{eff}}&=\frac{1}{2}\,{\sigma}^{{\frac
{4{\alpha}^{2}}{3K_{{1,1}}}}}\,{\delta}^{-{\frac {2K_{{1,-1}}}{K_{{1,1}}
}}} \left( {c}^{2}\rho+P \left( r \right)  \right)  \Biggl[ -{\frac {
\sigma\,\delta\,K_{{1,1}}P \left( r \right) \kappa}{\tau}}\\&-{\frac {K_{{1,1}
}}{\sigma\,\delta\,\tau\,{Y }^{{\alpha}^{2}}}}
+{\frac {\sigma\,\delta
\,K_{{1,1}}\Lambda\,{Y }^{{\alpha}^{2}}}{\tau}}+{\frac {\sigma\,\delta
\,K_{{1,1}}{\alpha}^{2}Y }{\tau\,{b}^{2}K_{{1,-1}}}}+{\frac {K_{{1,-3}
}}{\sigma\,\delta\,K_{{1,1}}}} \Biggr]
 \label{eq:dhee}
\end{split}
\end{equation}
with
$\tau={\frac {{K}_{1,1}^2}{{\Upsilon}^{{\alpha}^{2}}K_{{1,-1}}}}-{
\frac {{K}_{1,1}^2{\delta}^{2}{\sigma}^{2}\Upsilon\,{\alpha}^{2}}{
{{K}_{1,-1}^2}{b}^{2}}}-{\frac {{K}_{1,1}^2{\delta}^{2}{
\sigma}^{2}\Lambda\,{\Upsilon}^{{\alpha}^{2}}}{3\,K_{{1,-1}}}}-{
\frac {\kappa{c}^{2}\,K_{{1,1}}\,M_{{{\it eff}}}}{4\,\pi\left(  \left( \sigma\,
\delta \right) ^{{\frac {K_{{1,-3}}}{K_{{1,1}}}}} \right)}}
$\ ,

$\Upsilon= {\left( {\frac {b}{\sigma \delta }} \right)} ^{\frac{2}{K_{{1,1}}}}$, \ \ \
$\delta=R_{eff}^{\left(\frac{3K_{1,1}}{K_{3, -1}}\right)}$,\ \ \ $\sigma=\left(\frac{K_{3, -1}}{3K_{1,1}} \right)^{\left(\frac{K_{1,1}}{K_{3, -1}}\right)}$. \\
It is important to note that in dilaton gravity, gravitational mass, $m(r)=\int4\pi r^2\rho(r) dr$, and radius, $r$, are modified to $M_{eff}$ and $R_{eff}$ respectively.

\section{Dilaton HEE as a correction of TOV }\label{sec:Correct}
In this section, we use the same techniques as in the Ref.~\citealt{hendi2015dilatonic}. We expand Eq.~\ref{eq:dhee} in series of $\alpha$. When $\alpha$ has very small values, we can neglect all terms of order higher than $2$, then we have
\begin{equation}
 \begin{split}
  {\frac {dP}{dr}}=&{\frac { \left( 4\,{r}^{3}P(r) \kappa\,\pi-\frac{8}{3}\,\Lambda\,\pi\,{r}^{3}+\kappa\,{c}^{2}m(r)  \right)  \left( \rho{c}^{2}+P(r)
 \right) }{2\,r \left( \kappa\,{c}^{2}m(r)+\frac{4}{3}\,r\pi\,
 \bigl( \Lambda\,{r}^{2}-3 \bigr)  \right) }}\\
 &-{\alpha}^{2}\frac {72\, \bigl( \rho{c
}^{2}+P \left( r \right)  \bigr) }{r \bigl( 4\,
\Lambda\,\pi\,{r}^{3}+3\,\kappa\,{c}^{2}m \left( r \right) -12\,r\pi
 \bigr) ^{2}} \Biggl\{ \frac{\pi r}{3} \ln  \Bigl( {\frac {{b}^{2}}{{r}^{2}}} \Bigr)\\
 &\times\biggl[\frac{ 3}{4}\,\kappa\,{c}^{2} \left(\Lambda\,{r}^{2}+1 \right) m(r)
+ \pi \, \bigl[ \kappa\,\bigl( \Lambda\,{r}^{2}+3 \bigr) P(r) -4\,\Lambda
 \bigr] {r}^{3} \biggr]\\
&+\frac{1}{4}\,{\kappa}^{2} \left( m(r)  \right) ^{2}{c}^{4}+\pi\,{c}^{2} \kappa\,m(r)r \biggl[ \kappa\,P(r) \ln  \left( r \right) {r}^{2}+ \bigl( -\Lambda\,{r}^{2}\\
&+1 \bigr) \ln(r) +\frac{2}{3}\,\Lambda\,{r}^{2}-\frac{7}{4} \biggr]+{\pi}^{2}r \biggl[{r}^{2}\kappa\,P(r)\Bigl(\kappa\,{c}^{2}N(r)+\frac{2}{3}\Lambda\,{r}^{3}\\
&-r\Bigr) -\kappa\,{c}^{2}N(r) \bigl( \Lambda\,{r}^{2}-1 \bigr)-\frac{2}{9}\,{r}^{5}{\Lambda}^{2}-\frac{2}{3}\,\Lambda\,{r}^{3}+2\,r \biggr]\Biggr\}  \label{eq:dheeexp}
 \end{split}
\end{equation}
in which
$N(r) =\int \!-4\,{r}^{2}\ln  \left( r \right) \rho
 \left( r \right) \,{\rm d}r$.
 As it can be seen, the first term on the right-hand side of this equation is the well-known TOV equation, and the second term is considered as a correction term in dilaton gravity. A close inspection of Eq.~\ref{eq:dheeexp}, in the case of $\alpha=0$, the new HEE equation will be exactly the same as TOV.

 \section{Structure of Strange quark stars in dilaton gravity}\label{sec:structure}

In this section, we calculate some structural properties of strange quark stars(SQS) in dilaton gravity. In order to obtain the configurational characteristics of the quark stars, we must solve the HEE equation, Eq.~\ref{eq:dheeexp}, numerically by using an EoS of the form $P=P(\rho)$.
\subsection{EoS of strange quark matter}
\begin{figure}[ht]
\includegraphics[width=\columnwidth]{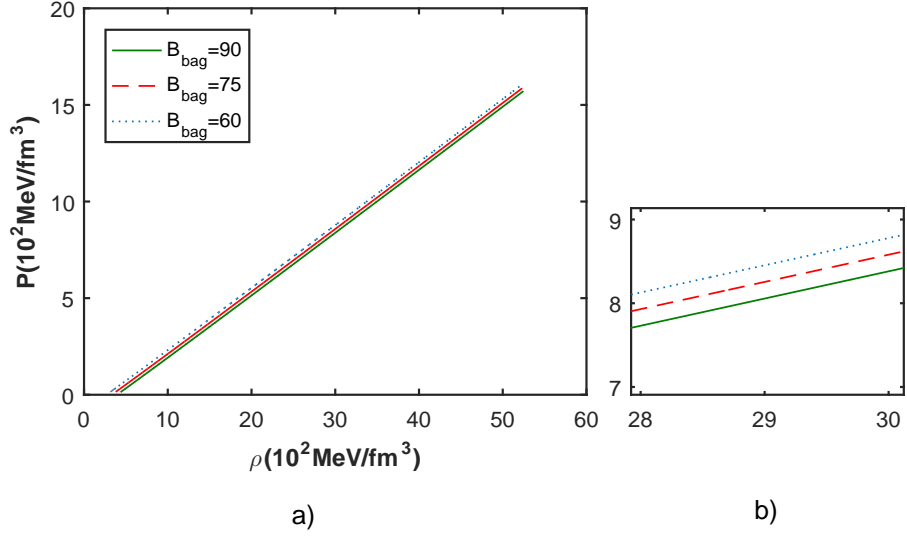}
\caption{EoS of SQM for various values of Bag constant(left panel). The right panel shows a zoom of the left one.} \label{fig:ff1}
\end{figure}

We obtain EoS of strange quark matter (SQM) using the MIT bag model. In this model, the total energy is the sum of kinetic
energy of quarks plus a bag constant ($B_{bag}$) ~(\citealt{chodos1974new}). In
fact, the bag constant $B_{bag}$ can be interpreted as the
difference between the energy densities of the noninteracting
quarks and the interacting ones. Dynamically it acts as a pressure,
that keeps the quark gas in constant density and potential. The value of
bag constant, $B_{bag}$, lies in the interval $58.8\,\frac{MeV}{fm^{3}}  <\,B_{bag}\,< 91.2\,\frac{MeV}{fm^{3}}$ ~(\citealt{stergioulas2003rotating}) that in this work, the values $60$, $75$ and $90 \frac{MeV}{fm^{3}}$ are considered for the bag constant. In order to obtain the equation of state, we have neglected the mass of $u$ and $d$ quarks and considered the mass of $s$ quark to be $m_{s}=150\, MeV$. More details on the EoS of SQM can be found in the Ref.~(\citealt{bordbar2011computation}). In Fig. \ref{fig:ff1} we have plotted behavior of EoS of SQM in MIT Bag model for various values of $B_{bag}$. This figure shows that for all values of bag constant EoS of SQM has linear treatment. Also, it is observed that EoS for SQM with a lower value of $B_{bag}$ is stiffer than the higher case. In the next section, we calculate maximum mass and corresponding radius of SQS in dilaton gravity by using this EoS and Eq.~\ref{eq:dheeexp} (also Eq.~\ref{eq:dhee}).

\subsection{Maximum mass of SQS in dilaton gravity}
Like other compact objects, SQSs reach a limit gravitational mass known as the maximum mass.
Subsequently, we can obtain the maximum mass ($M_{max}$) of SQS in dilaton gravity in the states with various values of $\alpha$ and $\Lambda$, by numerically integrating of Eq.~\ref{eq:dheeexp} (or Eq.~\ref{eq:dhee}) and using EoS of SQM.

In Tab.~\ref{tab:tnew}, We have calculated and presented structural properties of SQS  for variation of $\alpha$ with the reported value of the cosmological constant ($\Lambda\leq10^{-56}\,cm^{-2}$) and for vanishing cosmological constant. As we can see these values of $\Lambda$ and $\alpha$ have not the significant impact on the structural properties of SQS.

 It is worth mentioning that finding the exact value of the cosmological constant is an open issue. Although, its reported value can be valid in large-scale structure, but when we encounter with a local scale structure in close to a massive object such as neutron stars or SQSs, the value of $\Lambda$ may be very different from its reported value for large-scale structure ~(\citealt{bordbar2016neutron}). Motivated by this reason, $\Lambda\leq 10^{-13} cm^{-2}$ has been applied to study of structural properties of compact object in the presence of cosmological constant by several authors ~(\citealt{hendi2015dilatonic,bordbar2016neutron,Zubairi2015yda}). Also, it is notable that some of the cosmological models, consider $\Lambda$ as a decreasing variable with time, that has been predicted by decaying vacuum energy~(\citealt{waga1993decaying,bohmer2005dynamical}).
 Therefore, considering the effect of such different values of the cosmological constant on structural properties of a compact object can be interesting. In the remainder of this section, we have used $\Lambda\leq10^{-13} cm^{-2}$ as well, to obtain bulk properties of SQSs in dilaton gravity.

\begin{figure}[ht]
\includegraphics[width=\columnwidth]{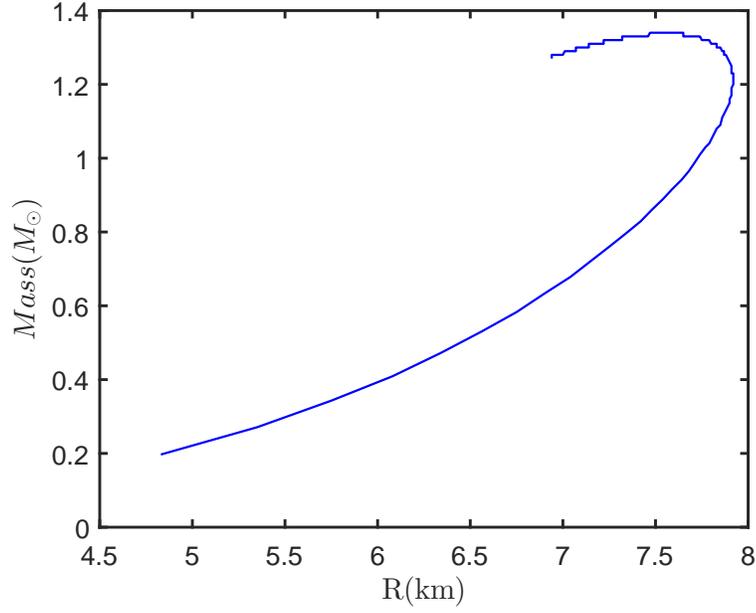}
\caption{The mass-radius relation of SQS for $\Lambda=0$ ($\alpha=0$ , $b=10^{-4}$ and $B_{bag}=90\,\frac{MeV}{fm^3}$) .} \label{fig:ff2}
\end{figure}

In Fig.~\ref{fig:ff2}, we have plotted the gravitational mass-radius for $\Lambda=\alpha=0$. The values of the maximum mass and the corresponding radius values are listed in~Tab.~\ref{tab:t1}. As we can see, these values are the same as $M_{max}$ and R in Einstein gravity ~(\citealt{bordbar2011computation}).

\begin{figure}[ht]
\includegraphics[width=\columnwidth]{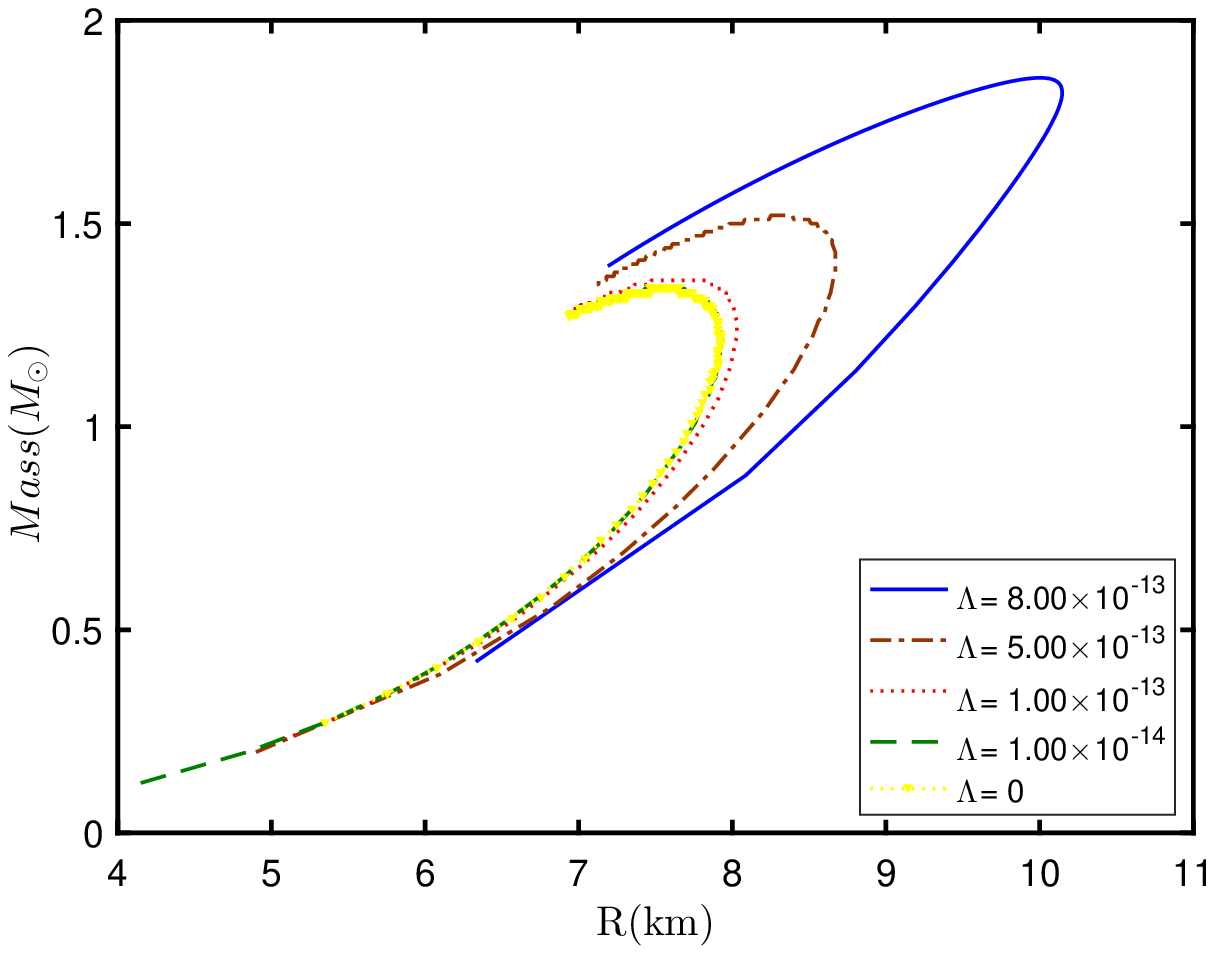}
\caption{The mass-radius relation of SQS for various positive values of
$\Lambda$ ($\alpha=0$, $b=10^{-4}$,$B_{bag}=90\,\frac{MeV}{fm^3}$) }\label{fig:ff3}
\end{figure}

\begin{figure}[ht]
\includegraphics[width=\columnwidth]{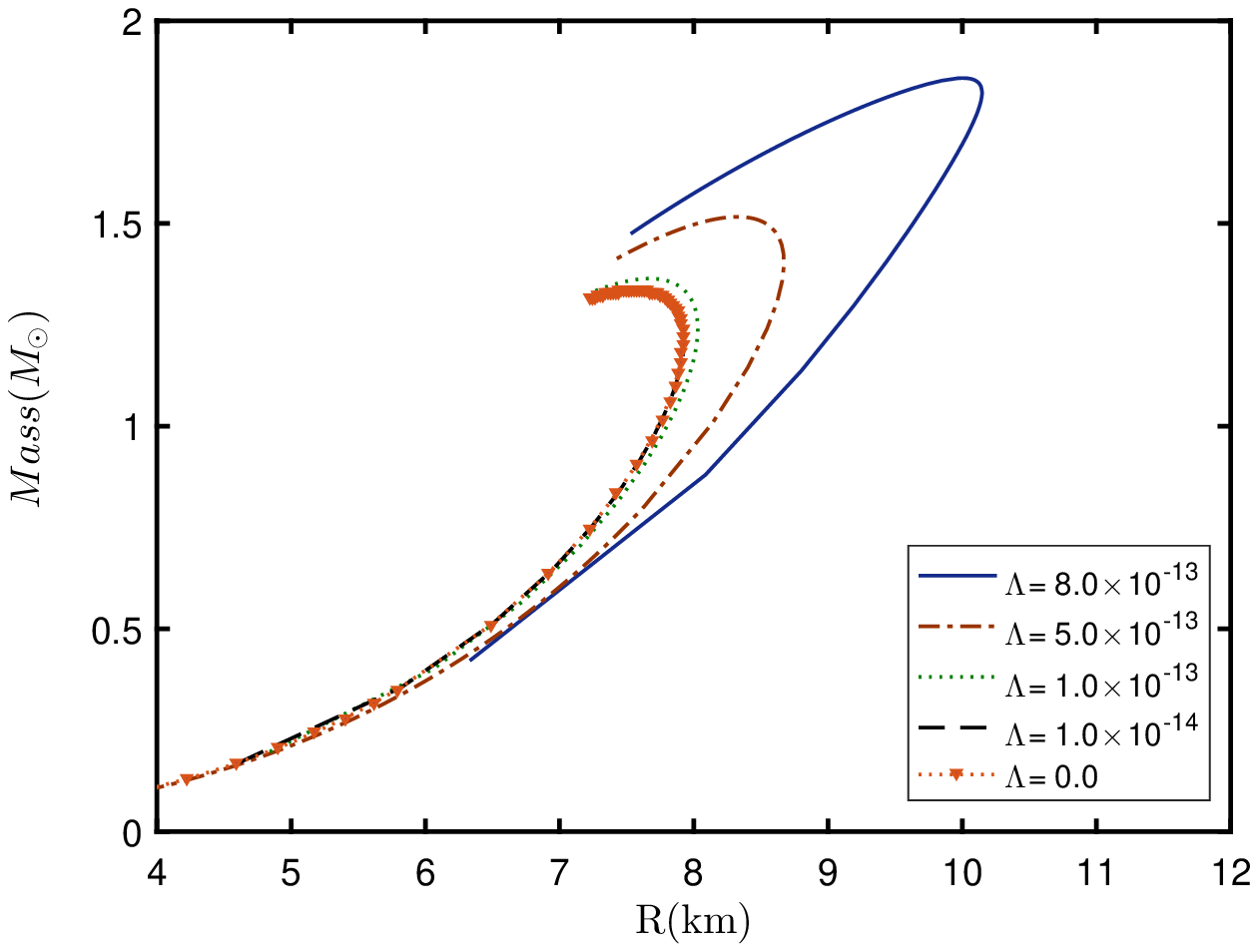}
\caption{The effective mass-effective radius relation of SQS for various positive values of
$\Lambda$ ($\alpha=0$, $b=10^{-4}$,$B_{bag}=90\,\frac{MeV}{fm^3}$) }\label{fig:ff4}
\end{figure}

\begin{figure}[ht]

\includegraphics[width=\columnwidth]{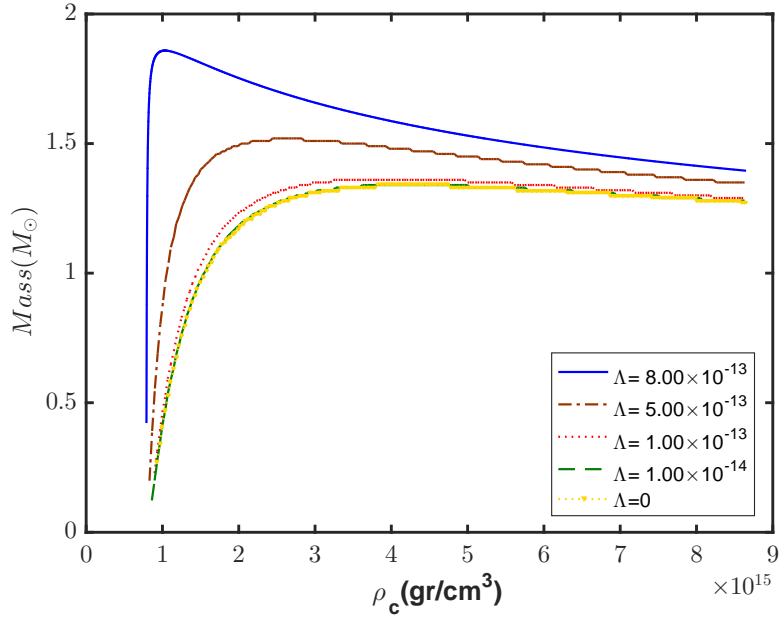}
\caption{The gravitational mass versus the total central energy density of the SQS with various values of
$\Lambda$ at $\alpha=0$, $b=10^{-4}$ and $B_{bag}=90\,\frac{Mev}{fm^{3}}$}\label{fig:ff5}
\end{figure}

The gravitational mass versus radius and central energy density for a constant value of $\alpha$ but in different values of $\Lambda$ is plotted in Figs.~\ref{fig:ff3} and \ref{fig:ff5}, respectively. We have also plotted the effective gravitational mass versus effective radius (see {Fig.~\ref{fig:ff4}}). Fig.~\ref{fig:ff3} shows that increasing in $\Lambda$ has led to an increase in the maximum mass. Moreover, for $\Lambda\leq10^{-14}$ the deviation of $M-R$ relation from Einstein gravity is very small (see Fig.~\ref{fig:ff3}). This result is repeated in Fig.~\ref{fig:ff4} and can be seen values of the gravitational mass and corresponding radius are same as effective mass and corresponding effective radius (see Tab.~\ref{tab:t1}). For a given central energy density ($\rho_c$), the mass of SQS increases by increasing cosmological constant (see Fig.~\ref{fig:ff5}). We can see that for higher values of the cosmological constant, increasing gravitational mass has a higher rate and reaching to $M_{max}$ occurs in SQSs with smaller values of central energy density ($\rho_c$).

\begin{figure}[ht]
\includegraphics[width=\columnwidth]{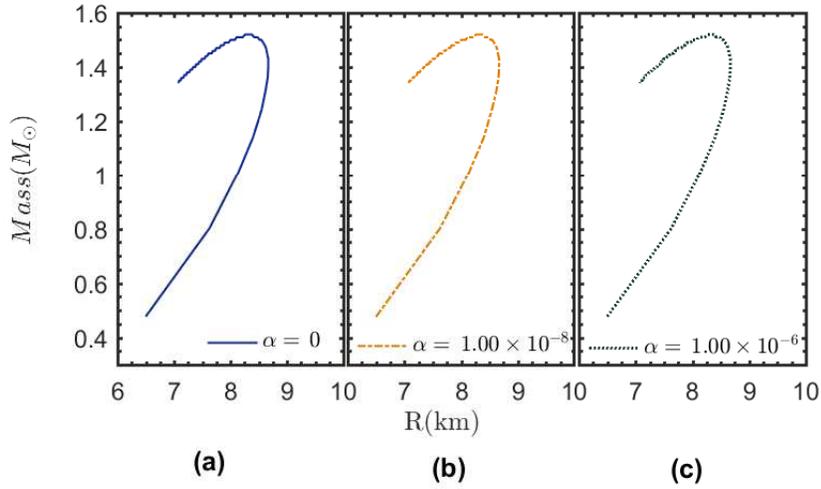}
\caption{The mass-radius relation of SQS for $\Lambda=5.00\times10^{-13}$ at \textbf{(a)}$\alpha=0$  \textbf{(b)}$\alpha=1.00\times10^{-8}$  \textbf{(c)}$\alpha=1.00\times10^{-6}$} \label{fig:ff6}
\end{figure}

In Fig.~\ref{fig:ff6}, we investigate the quark star structure for a given positive value of $\Lambda$ but for different values of $\alpha$, which shows that dilaton parameter, $\alpha$, variations do not affect in SQS structure. It should be noted that there is no answer to the diaton HEE for $\alpha\geq10^{-4}$.

\begin{figure}[ht]
\includegraphics[width=\columnwidth]{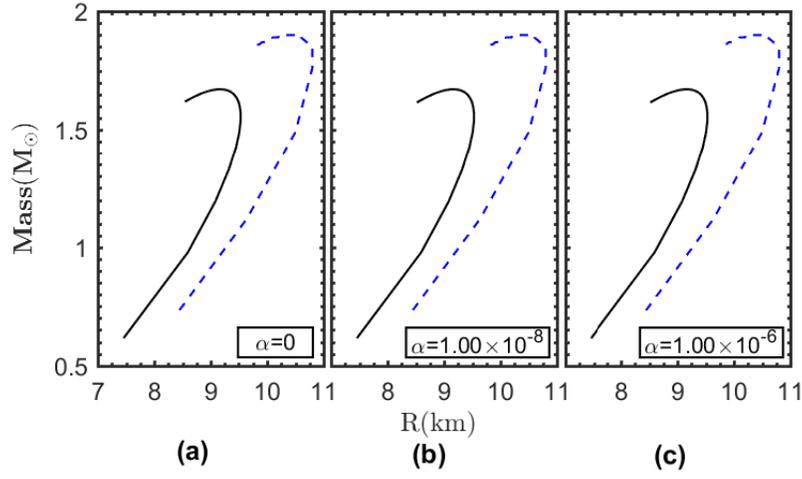}
\caption{As Fig.~\ref{fig:ff6},but for $B_{bag}=60\,\frac{MeV}{fm^3}$(dashed lines) and $B_{bag}=75\,\frac{MeV}{fm^3}$(solid lines) .} \label{fig:ff7}
\end{figure}

$B_{bag}=\,90\,\frac{MeV}{fm^{3}}$ is considered in figures drawn so far. In Fig.~\ref{fig:ff7}, we have plotted mass-radius relation of SQS in dilaton gravity for $B_{bag}=\,60$ and $75\,\frac{MeV}{fm^{3}}$ for a given value of $\Lambda$ and different values of $\alpha$. It can be seen that in these situations, the lack of influence of $\alpha$ on the mass and the radius of SQS is still remained. The values of the maximum mass and corresponding radius values in these modes are shown in Tabs.\ref{tab:t2} and \ref{tab:t3}.

\begin{figure}[ht]
\includegraphics[width=\columnwidth]{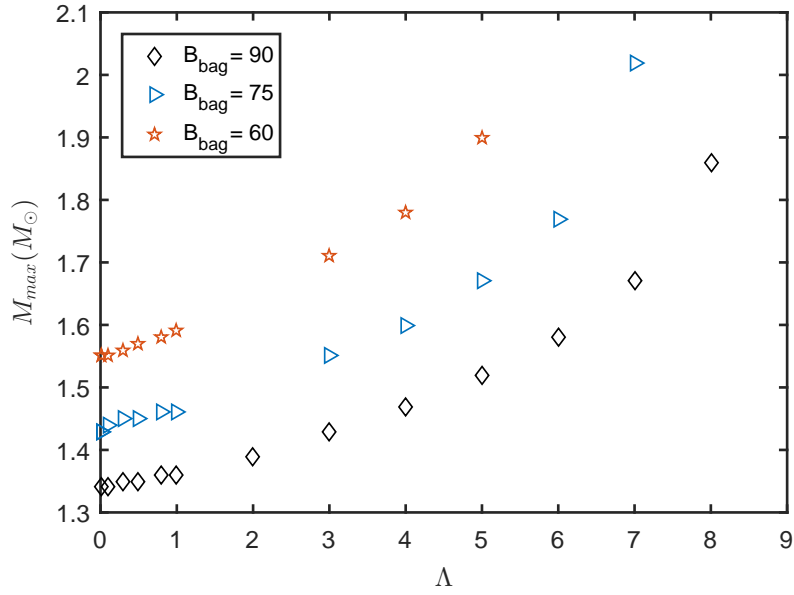}
\caption{Maximum mass of SQS versus cosmological constant ($\times10^{-13}$) for different bag constant with $\alpha=0$.} \label{fig:ff8}
\end{figure}

In Fig.~\ref{fig:ff8} the maximum gravitational mass as a function of $\Lambda$ in all cases of $B_{bag}$ is plotted. It shows that $M_{max}$ is an increasing function of $\Lambda$, however, the rate of the increase of $M_{max}$ versus $\Lambda$ increases with increasing bag constant.

\begin{figure}[ht]
\includegraphics[width=\columnwidth]{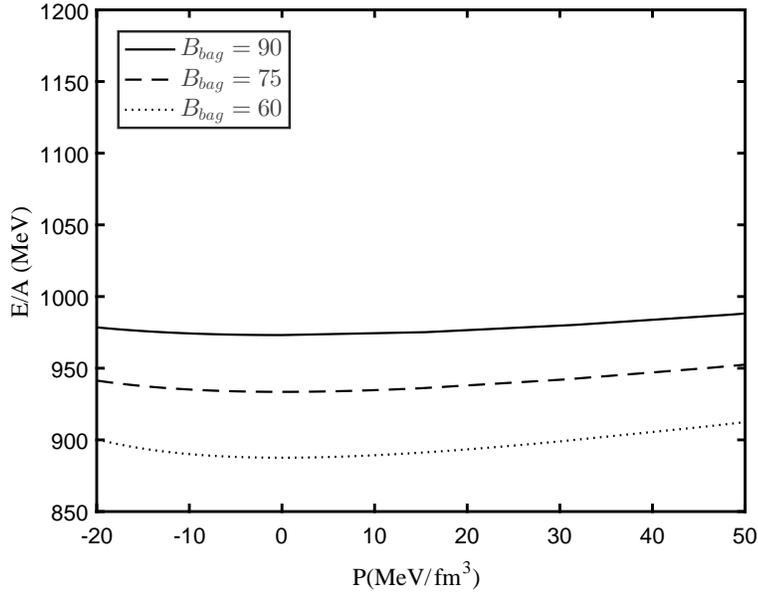}
\caption{The energy per baryon versus the pressure ($P$) for SQSs with the various values of bag constant.}\label{fig:ff9}
\end{figure}

In Fig.~\ref{fig:ff9}, we have plotted the energy per baryon ($E/A$) for SQM with the various values of 
Bag constant as a function of pressure. We see that the zero point of pressure for SQM considered with lower 
values of $B_{bag}$ has a lower E/A. This implies that SQMs considered with lower values of $B_{bag}$ are more stable than other ones.  %
In Tab.~\ref{tab:t4}, we have obtained the percentage increase of maximum mass of SQS and we have considered various values of the bag 
constant to calculate EoS of SQM in different values of $\Lambda$ with respect to obtained the $M_{max}$ in GR ($\alpha=\Lambda=0$) with same value of $B_{bag}$ (For instance ${(\Delta M)}^{B=90}_{\Lambda_{1}}=(M_{max})^{B=90}_{\Lambda_{1}=1\times10^{-13}}-(M_{max})^{B=90}_{\Lambda_{0}=0}=0.02\,
M_{sun}=1.49\%(M_{max})^{B=90}_{\Lambda_{0}=0}$ obtained for $B_{bag}=90$).
We can see that effect of $\Lambda$ on SQSs that have higher stability is upper than other ones(see columns of Tab.~\ref{tab:t4}). In addition, this table shows that increasing rate of the percentage increase of maximum mass of SQS, enhances by enhancing $\Lambda$ (see rows of Tab.~\ref{tab:t4}).

\begin{figure}[ht]
\includegraphics[width=\columnwidth]{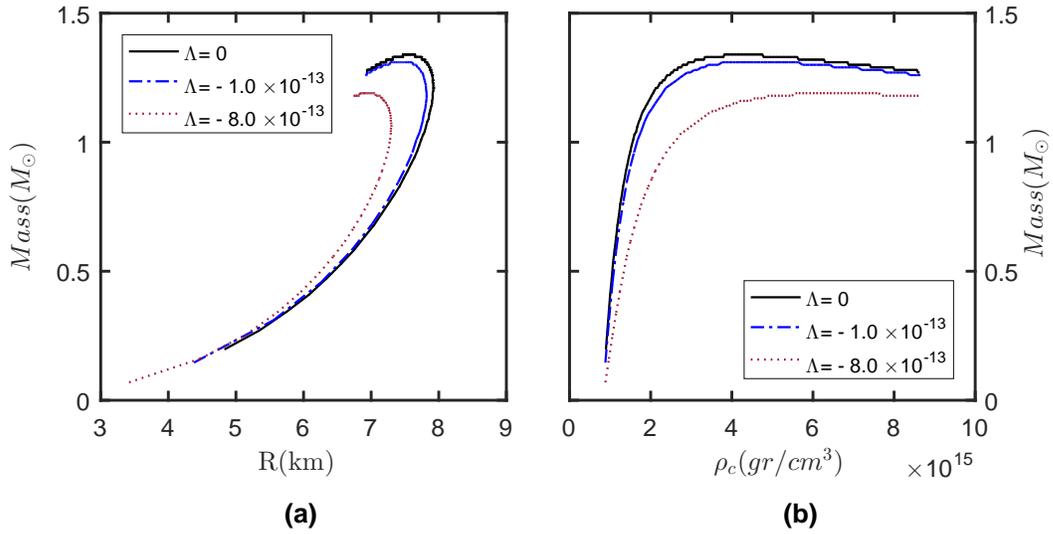}
\caption{The gravitational mass versus $\textbf{a)}$corresponding radius and $\textbf{b)}$central energy density($\rho_{c}$) of SQS in dilaton gravity for negative values of $\Lambda$($\alpha=0$ , $b=10^{-4}$ and $B_{bag}=90\,\frac{MeV}{fm^3}$).}\label{fig:ff10}
\end{figure}

Negative values of $\Lambda$ are investigated in Fig.~\ref{fig:ff10} in a particular value of $\alpha$ to obtain SQS structure. As we can see, increasing the maximum mass and its corresponding radius is simultaneous
with increasing in cosmological constant in negative values ( also see Tab.~\ref{tab:t5}).

It should be noted that for $\Lambda\geq10^{-12}$, HEE in dilaton gravity does not have a logical answer. The structural results for $\Lambda<10^{-14}$ are similar to those for $\Lambda=\,0$.
\subsection{Redshift and mean density}
The information about the compactness of a compact object, mass-radius ratio, can be found from the gravitational 
redshift that is an observational quantity(denoted by $z$). We can obtain $z$, by using Eq.~\ref{eq:ar} in two forms


\begin{equation}
 \begin{split}
z_{eff}=\Biggl\{\biggl[ -\,{\frac {\Lambda\,{(\Upsilon)}^{{\alpha}^{2}}}{3
K_{{1,-1}}}}-{\frac {\Upsilon \,{\alpha}^{2}}{{K^{2}_{{1,-1}}}{b}^{2}}}+{
\frac {({\Upsilon})^{-{\alpha}^{2}}}{{\Omega}^{2}K_{{1,-1}}}} \biggr] {\Omega}^{2}{K^{2}_{{1,1}}}\\ -\,{\frac {2\,G\,K_{{1,1}}M_{{{\it eff}}}\,{
\Omega}^{-{\frac {K_{{1,-3}}}{K_{{1,1}}}}}}{{c}^{2}}}\Biggr\}^{\! -\frac{1}{2}}-1
\end{split}
\label{eq:rsheff}
\end{equation}

and

\begin{equation}
\begin{split}
z=\Biggl\{\biggl[1-{\frac {2GM}{R{c}^{2}}}-\frac{\Lambda{R}^{2}}{3}\biggr]+\frac {{\alpha}^{2}}{3R{c}^{2}} \Biggl[ -R{c}^{2} \bigl( {R}^{2}
\Lambda+3 \bigr) \ln  \Bigl( {\frac {{b}^{2}}{{R}^{2}}} \Bigr) \\+
 \bigl( -3\Lambda\,{R}^{3}+6R \bigr) {c}^{2}-14GM \Biggr]\Biggr\}^{\! -\frac{1}{2}}-1
 \end{split}
 \label{eq:rsh}
\end{equation}
 where in Eq.~\ref{eq:rsheff}, $\Omega=\sigma\delta$ and $\Upsilon$ are functions of $R_{eff}$ and to obtain Eq.~\ref{eq:rsh}, we have expanded Eq.~\ref{eq:ar} in series of $\alpha$ and neglected all terms of order higher than $2$.

 In Tables \ref{tab:t1} to \ref{tab:t3} we have evaluated $z$ and $\overline{\rho}=\frac{3M}{4\pi R^3}$ (the mean energy density) for SQS in different cases. For all cases, we can see that with increasing $\Lambda$, gravitational redshift increases while $\overline{\rho}$ decreases (see Tabs. \ref{tab:t1} and \ref{tab:t3}). We can see that increasing in $\Lambda$ is equivalent to increasing $B_{bag}$. Therefore, it can be interpreted as increasing the values of the difference between the energy densities of non-interacting quarks and interacting ones. In addition, we can see that for fixed values of $\Lambda$, variations of $\alpha$ do not affect on $z$ and $\overline{\rho}$.

 \subsection{Energy conditions}
\begin{figure}[ht]
\includegraphics[width=\columnwidth]{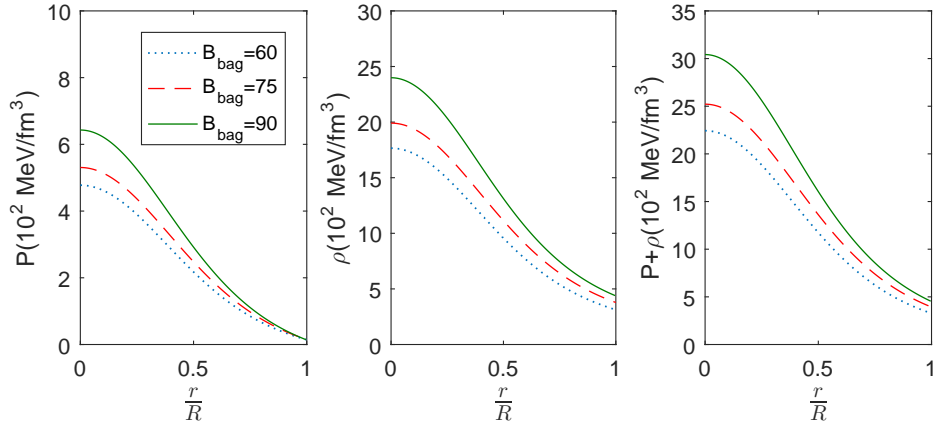}
\caption{The behavior of pressure $P$, energy density $\rho$ and $P+\rho$ with  respect to fractional radius $\frac{r}{R}$ under various values of $B_{bag}$ with $\alpha=\Lambda=0$.}\label{fig:ff11}
\end{figure}

\begin{figure}[ht]
\includegraphics[width=\columnwidth]{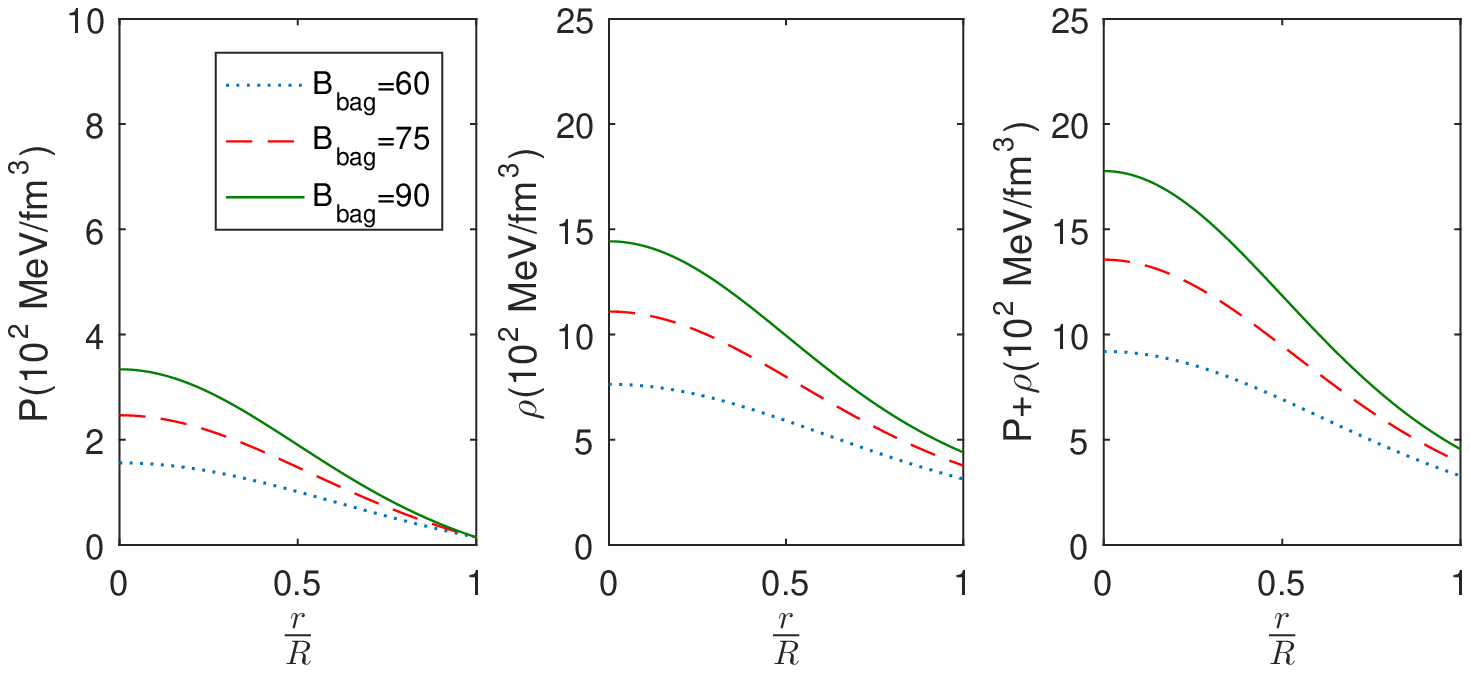}
\caption{As Fig. \ref{fig:ff11} for $\alpha=1\times10^{-6}$ and $\Lambda=5\times10^{-13}$.}\label{fig:ff12}
\end{figure}

Any acceptable physical model of isotropic fluid must satisfy conditions of energy (\citealt{hendi2016modified,visser2000energy}). The variation of pressure and energy density of a SQS with respect to fractional radius are shown in Figs.~\ref{fig:ff11} and ~\ref{fig:ff12}. As we can see the pressure of system is non-zero and positive at origin. Also, it is clear that the pressure and energy density decrease monotonically by increasing fractional radius. Moreover, It is observed that pressure reaches to zero at the boundary of the star ($r=R$) in all cases. Also, sum of pressure and energy density versus $\frac{r}{R}$ throughout of star are seen in these figures. It is clear that all energy conditions for a perfect fluidare satisfied for SQS in dilaton gravity.

 \subsection{Matching interior and exterior solutions}
 In this part, we investigate interior and exterior solutions matching at the boundaries by noting that we have assumed $\rho=\overline{\rho}$ throughout of star.
 At first, we can find the interior radial component of metric tensor ($\frac{{1}}{A(r)}$) by expanding Eq.~\ref{eq:ar} in series of $\alpha$ and neglecting all terms of order higher than 2, therefore for $r<R$ we have

 \begin{equation}\label{eq:arin}
 \begin{split}
  A(r)_{in}=\biggl[1-\frac{\Lambda\,{r}^{2}}{3}-\,{\frac {\kappa{c}^{2}m \left( r \right) }{4\,\pi\,r
}}\biggr]+ \biggl[2- \frac{1}{3}\,\Lambda\,{r}^{2}\ln  \left( {\frac {{b}^{2}}{{r}^{
2}}} \right) -\Lambda\,{r}^{2}-\ln  \left( {\frac {{b}^{2}}{{r}^{2}}
} \right) -\,{\frac {\kappa{c}^{2}m \left( r \right) }{4\,\pi\,r}}\\
+ {\frac{4}{3}\,\kappa\overline{\rho}\,{c}^{2}{r}^{2}\ln  \left( r \right) -\frac{4}{9}\,\kappa\overline{\rho}\,{c}^{2}{r}^
{2}}-{\frac {\kappa{c}^{2}m \left( r \right) \ln  \left( r \right) }{\pi\,r
}} \biggr] {\alpha}^{2}
\end{split}
 \end{equation}
 that by replacing $m(r)=\frac{M}{R^{3}}\,r^3$ and $\overline{\rho}=\frac{3\,M}{4\pi R^{3}}$ gets

 \begin{equation}\label{eq:arinshort}
 \begin{split}
 A(r)_{in}=\biggl[1-\frac{\Lambda\,{r}^{2}}{3}-\,{\frac {\kappa{c}^{2}M}{4\,\pi\,R^{3}
}}\,r^{2}\biggr]+ \biggl[2- \frac{1}{3}\,\Lambda\,{r}^{2}\ln  \left( {\frac {{b}^{2}}{{r}^{
2}}} \right) -\Lambda\,{r}^{2}-\ln  \left( {\frac {{b}^{2}}{{r}^{2}}
} \right)-{\frac {7\,\kappa{c}^{2}M}{12\,\pi\,{R}^{3}}}\,{r}^{2} \biggr] {\alpha}^{2}.
\end{split}
 \end{equation}

We can calculate time component ($B(r)$) of internal metric tensor by integrating the Eq.~\ref{eq:cemt} from specific value of r ($r<R$) to surface of star ($r=R$). We obtain


\begin{equation}\label{eq:bin}
{B \left( r \right)}_{in}=B \left( R \right)\left({\frac {\overline{\rho}\,{c}^{2}}{P(r)+\overline{\rho}\,{c
}^{2}}}\right)^2
\end{equation}
where $B(R)$ is the time component of metric at surface and $P(r)$ is pressure at specified radius that can be determined by using numerical integrating the Eq.~\ref{eq:dheeexp}. It is notable that if one insert $r=R$ in Eq.~\ref{eq:bin} ($P(r=R)=0$), ${B(r)}_{in}=B(R)$.

\begin{figure}[ht]
\includegraphics[width=\columnwidth]{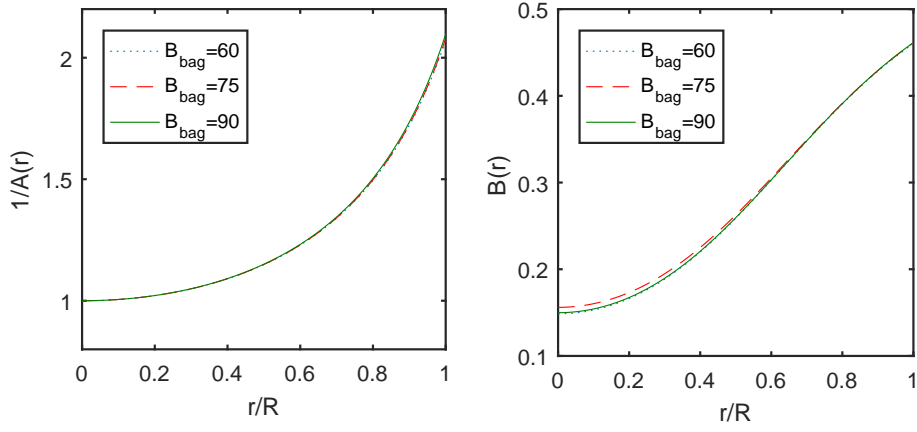}
\caption{The variation in radial and time component of metric tensor with  respect to fractional radius $\frac{r}{R}$ under various values of $B_{bag}$ with $\alpha=\Lambda=0$.}\label{fig:ff13}
\end{figure}

\begin{figure}[ht]
\includegraphics[width=\columnwidth]{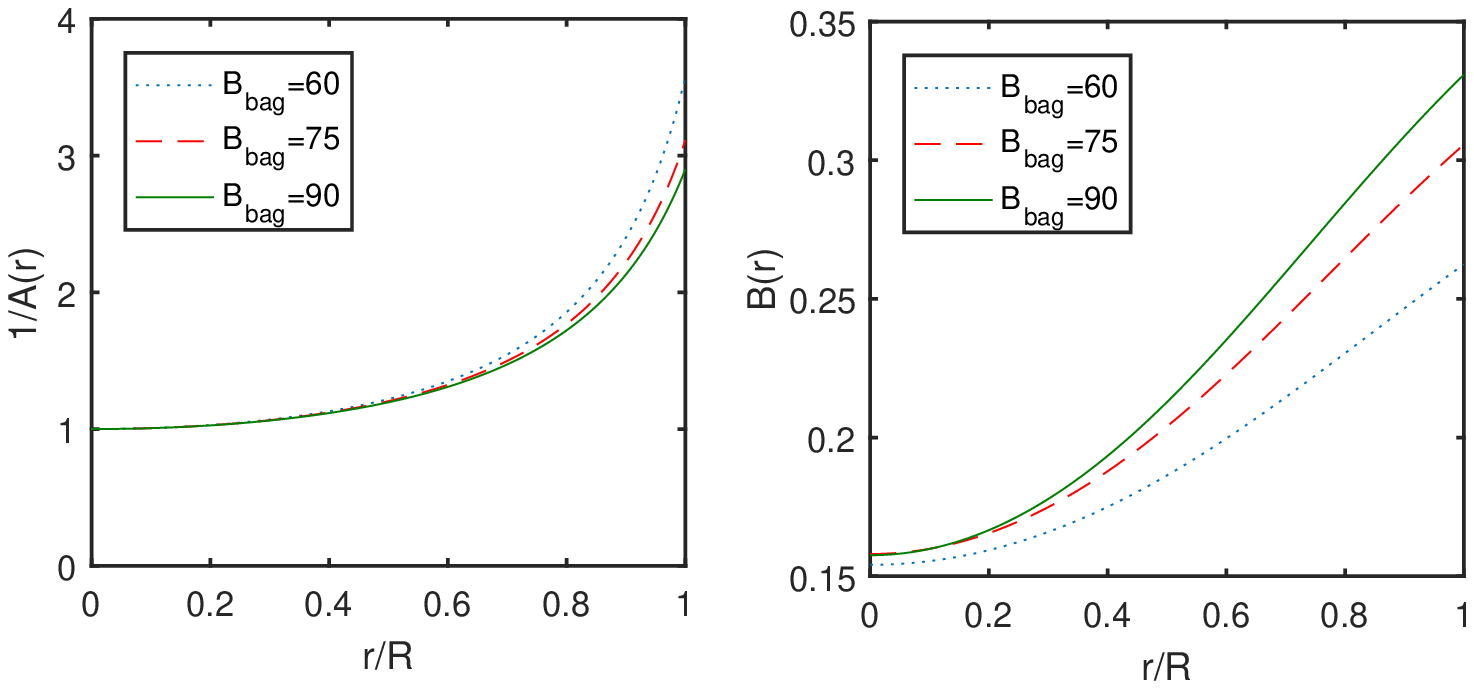}
\caption{As Fig.~\ref{fig:ff13} but for $\alpha=1\times10^{-6}$ and $\Lambda=5\times10^{-13}$. }\label{fig:ff14}
\end{figure}

We can obtain exterior radial and time component of metric, by solving Eqs.~\ref{eq:comp1} and \ref{eq:comp2} with considering that out of star, $P(r)=\rho(r)=0$. Then for $r>R$, we have

\begin{equation}\label{eq:about}
\begin{split}
A(r)_{out}=B(r)_{out}=\biggl[1-\,\frac{\Lambda\,{r}^{2}}{3}-\,{\frac {\kappa{c}^{2}M}{4\pi\,r}}\biggr]+ \biggl[2 -\frac{\,
\Lambda\,{r}^{2}\ln  \left( {\frac {{b}^{2}}{{r}^{2}}} \right)}{3} -
\Lambda\,{r}^{2}-\ln  \left( {\frac {{b}^{2}}{{r}^{2}}} \right) \\
-{
\frac {7\,\kappa{c}^{2}M}{12\,\pi\,r}}+{\frac {\kappa{c}^{2}M}{\pi\,r}\ln  \left( {
\frac {R}{r}} \right) } \biggr] {\alpha}^{2}.
\end{split}
\end{equation}
According to Eqs. \ref{eq:arin}-\ref{eq:about} , we can see easily that interior solution matches smoothly to the exterior solution at the surface of the star($r=R$). In Figs.~\ref{fig:ff13} and \ref{fig:ff14}, we have plotted $\frac{1}{A(r)}$ and $B(r)$ against fractional radius ($\frac{r}{R}$) for different values of $\alpha$, $\Lambda$ and $B_{bag}$. We can see that the component of metric at the center of object are non-singular and positive~(\citealt{shee2018anisotropic}). Also these figures show that $\frac{1}{A(r)}$ and $B(r)$ are monotonic increasing function of $\frac{r}{R}$ throughout the star as proved by Lake~(\citealt{lake2003all}) for any
physically acceptable model.

 \section{stability}\label{sec:dyst}
 \subsection{sound velocity}
\begin{figure}[ht]
\includegraphics[width=\columnwidth]{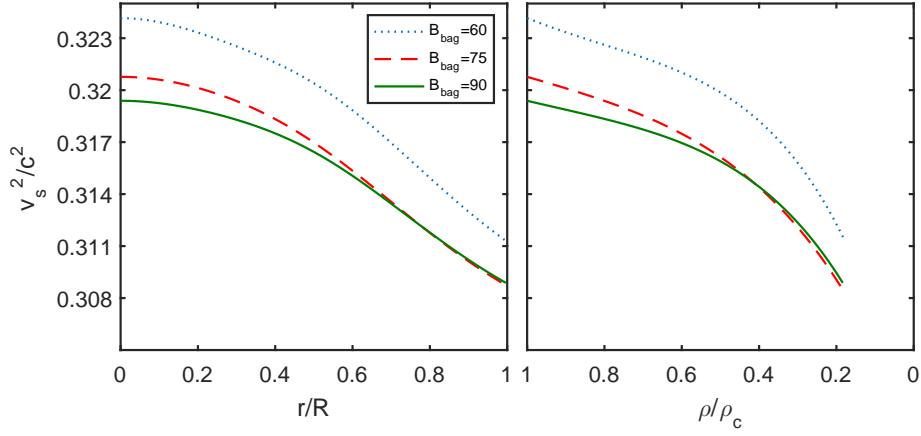}
\caption{The behavior of $\frac{v^{2}_{s}}{c^2}$ with  respect to fractional radius $\frac{r}{R}$ (left panel) and fractional energy density $\frac{\rho}{\rho_c}$ (right panel) under various values of $B_{bag}$ with $\alpha=\Lambda=0$.}\label{fig:ff15}
\end{figure}

\begin{figure}[ht]
\includegraphics[width=\columnwidth]{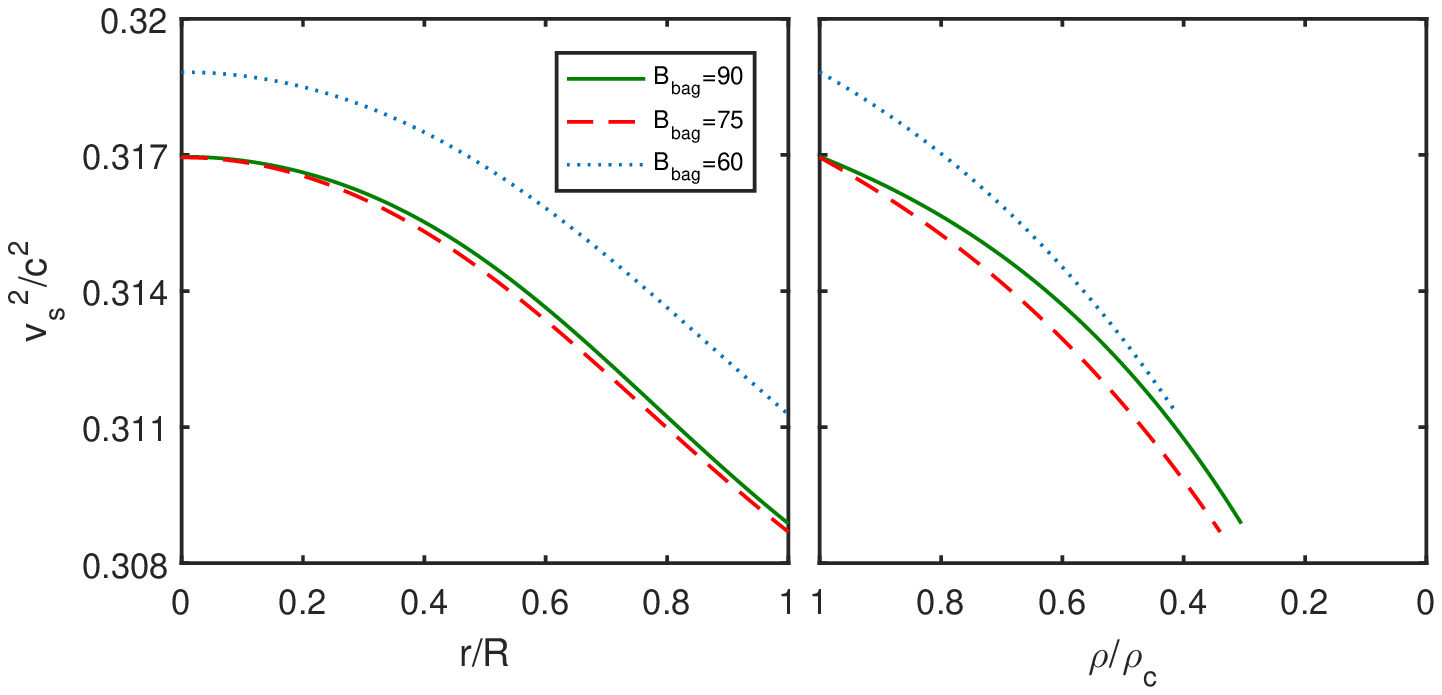}
\caption{As Fig.~\ref{fig:ff15} but for $\alpha=1\times10^{-6}$ and $\Lambda=5\times10^{-13}$.}\label{fig:ff16}
\end{figure}

 When we examine the stability of a compact object in terms of speed of sound, two conditions must be satisfied
\begin{enumerate}[label=\emph{(\roman*)}]
\item  \emph{The causality condition}: Speed of sound ($v_{s}$) have to be positive and less than light velocity ($c$). The variation of $\frac{v^{2}_{s}}{c^2}$ with the fractional radius and energy density ($\frac{\rho}{\rho_c}$) are shown in Figs.~\ref{fig:ff15} and \ref{fig:ff16}. It is observed that sound speed lies between 0 and 1 inside the quark star in all cases. Also $v_{s}$ decreases with increasing (decreasing) the radius (energy density) of star and reach to its minimum value at the surface of star.
\item  \emph{Herrera cracking concept}: The cracking concept proposed by Herrera~(\citealt{herrera1992cracking}) that has been applied to investigating the stability of an \emph{anisotropic} compact object (\citealt{shee2018anisotropic,deb2017relativistic,MAURYA2018152}). We do not need to check cracking concept, by noting that the model that we have considered here is the isotropic quark star that pressure of star is equal in all directions(see Eq.~\ref{eq:emtdiag}).
\end{enumerate}

 \subsection{Adiabatic index}
  An essential criterion for the dynamical stability of a compact object against radial adiabatic infinitesimal perturbations, is determined by adiabatic index $\Gamma$
  \begin{equation}
 \Gamma=\frac{\rho}{P} \bigl(1+{\frac{P}{\rho\,{c^{2}}}}\bigr) \frac{dP}{d\rho}.
\label{eq:adindex}
  \end{equation}

\begin{figure}[ht]
\includegraphics[width=\columnwidth]{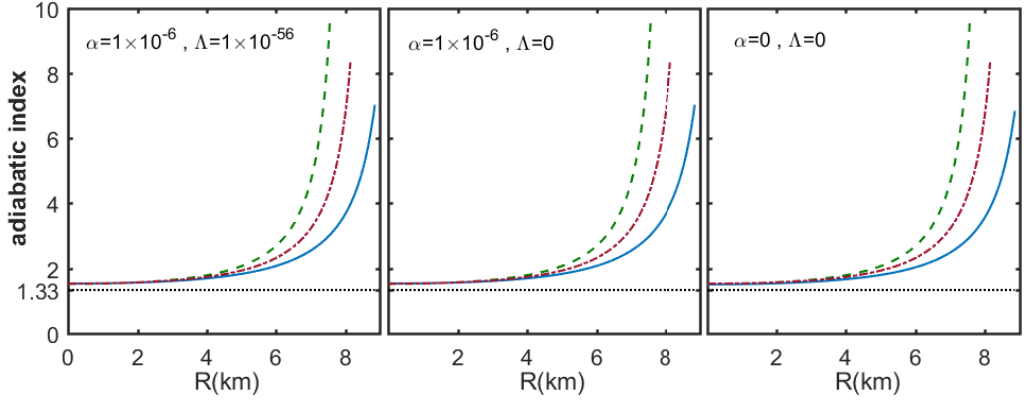}
\caption{The variation in adiabatic index with radial coordinate $r$  for different cases of $\alpha$ and $\Lambda$  with $B_{bag}=60\,\frac{MeV}{fm^3}$(solid line),$B_{bag}=75\,\frac{MeV}{fm^3}$(Dash-dot line) and $B_{bag}=90\,\frac{MeV}{fm^3}$(dashed line).}\label{fig:ff17}
\end{figure}

\begin{figure}[ht]
\includegraphics[width=\columnwidth]{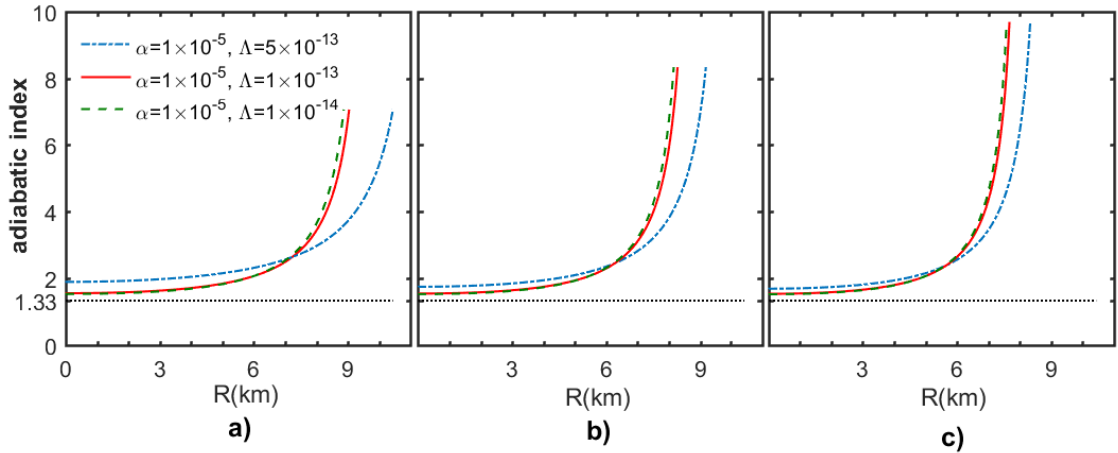}
\caption{The variation in adiabatic index with radial coordinate r under various condition of $\alpha$ and $\Lambda$ in\textbf{(a)}$B_{bag}=60\,\frac{MeV}{fm^3}$, \textbf{(b)}$B_{bag}=75\,\frac{MeV}{fm^3}$ and \textbf{(c)}$B_{bag}=90\,\frac{MeV}{fm^3}$}\label{fig:ff18}
\end{figure}

   Chandrasekhar showed that for the dynamical stability of a compact object, $\Gamma$ must be greater than $\frac{4}{3}$ throughout it (\citealt{chandrasekhar1964dynamical}). The variation of $\Gamma$ versus radial coordinate $r$ is shown in Figs.~\ref{fig:ff17} and~\ref{fig:ff18} for different cases of $\alpha$, $\Lambda$ and Bag constant. One can see that the variation of $\alpha$ has no effect on the adiabatic index values while the variation of $\Lambda$ increases the rate of increasing $\Gamma$. In amounts other than the reported value of $\Lambda$, adiabatic index reaches to the greater values near the surface of the star in larger values of Bag constant (see Fig.~\ref{fig:ff18}). Also, we can see in Figs.~\ref{fig:ff17} and~\ref{fig:ff18}, the values of $\Gamma$ for central layers of SQS star is closer to and greater than $\frac{4}{3}$ but for layers near the surface are very large. In all cases that we have considered here, the values of $\Gamma$ are higher than $\frac{4}{3}$.

 Therefore, we can conclude SQS in dilaton gravity is stable in all values of bag constant that we have considered due to causality condition and adiabatic index inequality ($\Gamma\geq\frac{4}{3}$) are valid everywhere inside the star.


\section{Conclusion}\label{sec:con}
We have obtained HEE of a compact object in dilaton gravity by using two approaches and then by using the obtained HEEs we have calculated some structural properties of SQS. We have assumed that the dilaton gravity has constructed from dilaton field with a potential, including two Liouville type terms in the background of Einstein gravity. We have seen that in the values of $\alpha$ and $\Lambda$ where HEE in dilaton gravity has a logical answer, SQS does not change with variations of $\alpha$. This behavior also persists in different values of the bag constant. On the other hand, increasing $\Lambda$ enhances the maximum mass. We have seen that the percentage increase of maximum mass of SQS has higher values for SQSs that are more stable than others. %
We have shown that effect of dilaton gravity and applying smaller bag constant in EoS of SQM on the SQS has led to existence SQSs with bigger masses and radii that are more stable than SQSs in Einstein gravity.
Also, we have shown that SQSs in dilaton gravity are stable against radial adiabatic infinitesimal perturbations in all cases that we have considered here. Moreover, our result show causality condition and energy conditions are valid for the model that has considered.
 Since the values of $M_{max}$ and radius for $\Lambda<\,10^{-14}$ are the same values for $\Lambda=0$ and cosmological observations suggest $\Lambda\leq3\times10^{-56}{cm}^{-2}$, it can be concluded that this limit of the cosmological constant does not affect on SQS structure in dilaton gravity.
%
%
\normalem
\begin{acknowledgements}
{G. H. Bordbar wishes to thank Shiraz University
Research Council. This work has been supported by Research Institute for Astronomy
and Astrophysics of Maragha. }
\end{acknowledgements}


\begin{thebibliography}{55}
\providecommand\natexlab[1]{#1}
\providecommand\JournalTitle[1]{#1}

\bibitem[Alcock {et~al.}(1986{\natexlab{a}})]{alcock1986model}
Alcock, C., Farhi, E., \& Olinto, A. 1986{\natexlab{a}}, Physical Review
  Letters, 57, 2088

\bibitem[Alcock {et~al.}(1986{\natexlab{b}})]{alcock1986strange}
Alcock, C., Farhi, E., \& Olinto, A. 1986{\natexlab{b}}, The Astrophysical
  Journal, 310, 261

\bibitem[Alcock \& Olinto(1988)]{alcock1988exotic}
Alcock, C., \& Olinto, A. 1988, Annual Review of Nuclear and Particle Science,
  38, 161

\bibitem[{Arapo{\v g}lu} {et~al.}(2011)]{arapouglu2011constraints}
{Arapo{\v g}lu}, S., {Deliduman}, C., \& {Ek{\c s}i}, K.~Y. 2011, Journal of Cosmology and Astroparticle Physics, 7, 020

\bibitem[Astashenok(2016)]{astashenok2016neutron}
Astashenok, A. 2016, in International Journal of Modern Physics: Conference
  Series, Vol.~41, World Scientific, 1660130

\bibitem[Astashenok {et~al.}(2015)]{astashenok2015nonperturbative}
Astashenok, A., Capozziello, S., \& Odintsov, S.~D. 2015, Physics Letters B,
  742, 160

\bibitem[{Astashenok} {et~al.}(2013)]{astashenok2013further}
{Astashenok}, A.~V., {Capozziello}, S., \& {Odintsov}, S.~D. 2013, Journal of Cosmology and Astroparticle Physics, 12,
  040

\bibitem[{Balaguera-Antol{\'{\i}}nez}
  {et~al.}(2005)]{balaguera2005astrophysical}
{Balaguera-Antol{\'{\i}}nez}, A., {Nowakowski}, M., \& {B{\"o}hmer}, C.~G.
  2005, International Journal of Modern Physics D, 14, 1507

\bibitem[{B{\"o}hmer} \& {Harko}(2005{\natexlab{a}})]{bohmer2005does}
{B{\"o}hmer}, C.~G., \& {Harko}, T. 2005{\natexlab{a}}, Physics Letters B, 630,
  73

\bibitem[{B{\"o}hmer} \& {Harko}(2005{\natexlab{b}})]{bohmer2005dynamical}
{B{\"o}hmer}, C.~G., \& {Harko}, T. 2005{\natexlab{b}}, Physical Review D, 71, 084026

\bibitem[{Bombaci}(1997)]{bombaci1997observational}
{Bombaci}, I. 1997, \prc, 55, 1587

\bibitem[{Bordbar} {et~al.}(2006)]{bordbar2006structure}
{Bordbar}, G.~H., {Bigdeli}, M., \& {Yazdizadeh}, T. 2006, International
  Journal of Modern Physics A, 21, 5991

\bibitem[Bordbar {et~al.}(2016)]{bordbar2016neutron}
Bordbar, G.~H., Hendi, S.~H., \& Eslam~Panah, B. 2016, The European Physical
  Journal Plus, 131, 315

\bibitem[Bordbar \& Peivand(2011)]{bordbar2011computation}
Bordbar, G., \& Peivand, A.~R. 2011, Research in Astronomy and Astrophysics,
  11, 851

\bibitem[Chan {et~al.}(1995)]{chan1995charged}
Chan, K., Horne, J., \& Mann, R. 1995, Nuclear Physics B, 447, 441

\bibitem[Chandrasekhar(1964)]{chandrasekhar1964dynamical}
Chandrasekhar, S. 1964, Physical Review Letters, 12, 114

\bibitem[Cho(1990)]{cho1990unified}
Cho, Y. 1990, Physical Review D, 41, 2462

\bibitem[Chodos {et~al.}(1974)]{chodos1974new}
Chodos, A., Jaffe, R., Johnson, K., Thorn, C., \& Weisskopf, V. 1974, Physical
  Review D, 9, 3471

\bibitem[Deb {et~al.}(2017)]{deb2017relativistic}
Deb, D., Chowdhury, S., Ray, S., Rahaman, F., \& Guha, B. 2017, Annals of
  Physics, 387, 239

\bibitem[Dehghani \& Farhangkhah(2005)]{dehghani2005charged}
Dehghani, M., \& Farhangkhah, N. 2005, Physical Review D, 71, 044008

\bibitem[Demorest {et~al.}(2010)]{demorest2010two}
Demorest, P., Pennucci, T., Ransom, S., Roberts, M., \& Hessels, J. 2010,
  Nature, 467, 1081

\bibitem[Dey {et~al.}(1998)]{dey1998strange}
Dey, M., Bombaci, I., Dey, J., Ray, S., \& Samanta, B. 1998, Physics Letters B,
  438, 123

\bibitem[Farhi \& Jaffe(1984)]{farhi1984strange}
Farhi, E., \& Jaffe, R.~L. 1984, Physical Review D, 30, 2379

\bibitem[Gangopadhyay {et~al.}(2013)]{gangopadhyay2013strange}
Gangopadhyay, T., Ray, S., Li, X., Dey, J., \& Dey, M. 2013, Monthly Notices of
  the Royal Astronomical Society, 431, 3216

\bibitem[Glendenning(1990)]{glendenning1990nk}
Glendenning, N. 1990, Modern Physics Letters A, 5, 2197

\bibitem[G{\"u}ver {et~al.}(2010)]{guver2010mass}
G{\"u}ver, T., Wroblewski, P., Camarota, L., \& {\"O}zel, F. 2010, The
  Astrophysical Journal, 719, 1807

\bibitem[Haensel {et~al.}(1986)]{haensel1986strange}
Haensel, P., Zdunik, J., \& Schaefer, R. 1986, Astronomy and Astrophysics, 160,
  121

\bibitem[Hendi {et~al.}(2015)]{hendi2015dilatonic}
Hendi, S., Bordbar, G., Eslam~Panah, B., \& Najafi, M. 2015, Astrophysics and
  Space Science, 358, 30

\bibitem[Hendi {et~al.}(2017)]{hendi2017neutron}
Hendi, S., Bordbar, G., Eslam~panah, B., \& Panahiyan, S. 2017, Journal of
  Cosmology and Astroparticle Physics, 2017, 004

\bibitem[Hendi {et~al.}(2016)]{hendi2016charged}
Hendi, S.~H., Faizal, M., Eslam~Panah, B., \& Panahiyan, S. 2016, The European
  Physical Journal C, 76, 296

\bibitem[Herrera(1992)]{herrera1992cracking}
Herrera, L. 1992, Physics Letters A, 165, 206

\bibitem[Katsuragawa {et~al.}(2016)]{katsuragawa2016relativistic}
Katsuragawa, T., Nojiri, S., Odintsov, S., \& Yamazaki, M. 2016, Physical
  Review D, 93, 124013

\bibitem[Knop~et al.(2003)]{knop2003new}
Knop~et al., R. 2003, The Astrophysical Journal, 598, 102

\bibitem[Lake(2003)]{lake2003all}
Lake, K. 2003, Physical Review D, 67, 104015

\bibitem[Maurya {et~al.}(2018)]{MAURYA2018152}
Maurya, S., Ray, S., Ghosh, S., Manna, S., \& T.T., S. 2018, Annals of Physics,
  395, 152

\bibitem[Momeni \& Myrzakulov(2015)]{momeni2015tolman}
Momeni, D., \& Myrzakulov, R. 2015, International Journal of Geometric Methods
  in Modern Physics, 12, 1550014

\bibitem[Narain {et~al.}(2006)]{narain2006compact}
Narain, G., Schaffner-Bielich, J., \& Mishustin, I.~N. 2006, Physical Review D,
  74, 063003

\bibitem[Oppenheimer \& Volkoff(1939)]{oppenheimer1939massive}
Oppenheimer, J.~R., \& Volkoff, G.~M. 1939, Physical Review, 55, 374

\bibitem[Paul {et~al.}(2011)]{paul2011relativistic}
Paul, B., Chattopadhyay, P., Karmakar, S., \& Tikekar, R. 2011, Modern Physics
  Letters A, 26, 575

\bibitem[Perlmutter~et al.(1999)]{perlmutter1999measurements}
Perlmutter~et al., S. 1999, The Astrophysical Journal, 517, 565

\bibitem[Rawls {et~al.}(2011)]{rawls2011refined}
Rawls, M.~L., Orosz, J., McClintock, J., {et~al.} 2011, The Astrophysical
  Journal, 730, 25

\bibitem[Riess {et~al.}(1998)]{riess1998observational}
Riess, A., Filippenko, A., Challis, P., {et~al.} 1998, The Astronomical
  Journal, 116, 1009

\bibitem[S.H. {et~al.}(2016)]{hendi2016modified}
S.H., H., Panah, B. G. B.~E., \& Panahiyan, S. 2016, Journal of Cosmology and
  Astroparticle Physics, 2016, 013

\bibitem[Sharif \& Fatima(2016)]{sharif2016static}
Sharif, M., \& Fatima, H. 2016, International Journal of Modern Physics D, 25,
  1650083

\bibitem[Shee {et~al.}(2018)]{shee2018anisotropic}
Shee, D., Deb, D., Ghosh, S., Ray, S., \& Guha, B. 2018, International Journal
  of Modern Physics D, 1850089

\bibitem[Sheykhi {et~al.}(2006)]{sheykhi2006asymptotically}
Sheykhi, A., Riazi, N., \& Mahzoon, M. 2006, Physical Review D, 74, 044025

\bibitem[{Silbar} \& {Reddy}(2004)]{silbar2005erratum}
{Silbar}, R.~R., \& {Reddy}, S. 2004, American Journal of Physics, 72, 892

\bibitem[Stergioulas(2003)]{stergioulas2003rotating}
Stergioulas, N. 2003, Living Reviews in Relativity, 6, 3

\bibitem[Stuchl{\'\i}k(2000)]{stuchlik2000spherically}
Stuchl{\'\i}k, Z. 2000, ACTA PHYSICA SLOVACA, 50, 219

\bibitem[Tolman(1939)]{tolman1939static}
Tolman, R.~C. 1939, Physical Review, 55, 364

\bibitem[Tonry {et~al.}(2003)]{tonry2003cosmological}
Tonry, J., Schmidt, B., Barris, B., {et~al.} 2003, The Astrophysical Journal,
  594, 1

\bibitem[Visser \& Barcelo(2000)]{visser2000energy}
Visser, M., \& Barcelo, C. 2000, in Cosmo-99 (World Scientific), 98

\bibitem[Waga(1993)]{waga1993decaying}
Waga, I. 1993, The Astrophysical Journal, 414, 436

\bibitem[Witten(1984)]{witten1984witten}
Witten, E. 1984, Physical Review D, 30, 272

\bibitem[Zubairi {et~al.}(2015)]{Zubairi2015yda}
Zubairi, O., Romero, A., \& Weber, F. 2015, Journal of Physics: Conference Series, 615, 012003

\end{thebibliography}

\newpage
\begin{table}

\caption{   Maximum gravitational
mass($M_{max}$) and the corresponding radius($R$) of  SQS  in dilaton gravity for various positive values of $\Lambda$ and $\alpha$ with various values of $B_{bag}$.}\label{tab:tnew}

\begin{tabular}{lcccccr}
\hline
${B_{bag}(\frac{MeV}{fm^{3}})}$ & ${\alpha}$  &  ${\Lambda}$  &  ${M_{max}(M_{\odot})}$ & ${R(Km)}$  &  ${\overline{\rho}}(10^{15}\frac{gr}{cm^{3}})$  &  ${z(10^{-1})}$  \\ \hline
\multirow{6}{*}{60} &0                   & 0                    &  1.55 & 8.86 & 1.06 & 4.37  \\
&$1.00\times10^{-8}$ & 0                    &  1.55 & 8.86 & 1.06 & 4.37  \\
&$1.00\times10^{-5}$ & 0                    &  1.55 & 8.86 & 1.06 & 4.37  \\
&0                   & $1.00\times10^{-56}$ &  1.55 & 8.86 & 1.06 & 4.37  \\
&$1.00\times10^{-8}$ & $1.00\times10^{-56}$ &  1.55 & 8.86 & 1.06 & 4.37  \\
&$1.00\times10^{-5}$ & $1.00\times10^{-56}$ &  1.55 & 8.86 & 1.06 & 4.37  \\
\hline
\multirow{6}{*}{75} &0                   & 0                    &  1.43 & 8.15 & 1.25 & 4.40  \\
&$1.00\times10^{-8}$ & 0                    &  1.43 & 8.15 & 1.25 & 4.40  \\
&$1.00\times10^{-5}$ & 0                    &  1.43 & 8.15 & 1.25 & 4.40  \\
&0                   & $1.00\times10^{-56}$ &  1.43 & 8.15 & 1.25 & 4.40  \\
&$1.00\times10^{-8}$ & $1.00\times10^{-56}$ &  1.43 & 8.15 & 1.25 & 4.40  \\
&$1.00\times10^{-5}$ & $1.00\times10^{-56}$ &  1.43 & 8.15 & 1.25 & 4.40  \\
\hline
\multirow{6}{*}{90} &0                   &0                     &  1.34 & 7.58 & 1.46 & 4.45  \\
&$1.00\times10^{-8}$ &0                     &  1.34 & 7.58 & 1.46 & 4.45  \\
&$1.00\times10^{-5}$ &0                     &  1.34 & 7.58 & 1.46 & 4.45  \\
&0                   & $1.00\times10^{-56}$ &  1.34 & 7.58 & 1.46 & 4.45  \\
&$1.00\times10^{-8}$ & $1.00\times10^{-56}$ &  1.34 & 7.58 & 1.46 & 4.45  \\
&$1.00\times10^{-5}$ & $1.00\times10^{-56}$ &  1.34 & 7.58 & 1.46 & 4.45  \\
\hline

\end{tabular}

\end{table}
\begin{table}[tbp]
\caption{\label{tab:t1} Maximum gravitational mass($M_{max}$) and the corresponding radius($R$) of SQS in dilaton gravity for various positive values of $\Lambda$ at $b=10^{-4}$ and $\alpha=0$}
\begin{tabular}{lccccccc}
\hline${\Lambda}$ & ${M_{max}(M_{\odot})}$ & ${R(Km)}$ & ${M^{eff}_{max}(M_{\odot})}$ & ${R^{eff}%
(Km)}$ & ${\overline{\rho}}(10^{15}gr \,cm^{-3})$ & ${z_{eff}(10^{-1})}$ & ${z}(10^{-1})$ \\
\hline
         0           & 1.34 & 7.58 & 1.34 & 7.58 & 1.46 & 4.45 & 4.45 \\
$1.00\times10^{-14}$ & 1.34 & 7.58 & 1.34 & 7.58 & 1.43 & 4.48 & 4.48 \\
$1.00\times10^{-13}$ & 1.36 & 7.68 & 1.36 & 7.68 & 1.43 & 4.78 & 4.78 \\
$3.00\times10^{-13}$ & 1.43 & 7.95 & 1.43 & 7.95 & 1.35 & 5.69 & 5.69 \\
$4.00\times10^{-13}$ & 1.47 & 8.13 & 1.47 & 8.13 & 1.30 & 6.25 & 6.25 \\
$5.00\times10^{-13}$ & 1.52 & 8.34 & 1.52 & 8.34 & 1.24 & 6.99 & 6.99  \\
$8.00\times10^{-13}$ & 1.86 & 10.01 & 1.86 & 10.01 & 0.88 & 13.27 & 13.27 \\
\hline

\end{tabular}
\end{table}


\begin{table}

\caption{ Maximum gravitational
mass($M_{max}$) and the corresponding radius($R$) of SQS in dilaton gravity for various positive values of $\Lambda$ and $\alpha$ with $B_{bag}=60\frac{MeV}{fm^{3}}$.}

\begin{tabular}{lcccccc}
\hline
${\alpha}$ & ${\Lambda}$ & ${M_{max}(M_{\odot})}$ & ${R
(Km)}$ & ${\overline{\rho}}(10^{15}\frac{gr}{cm^{3}})$ & ${z(10^{-1})}$ & ${B_{bag}(\frac{MeV}{{fm}^{3}})}$ \\ \hline

$1.00\times10^{-7}$         & $1.00\times10^{-15}$ &  1.55 & 8.86  & 1.06 & 4.38 & 60 \\
$1.00\times10^{-7}$         & $1.00\times10^{-14}$ &  1.55 & 8.88  & 1.06 & 4.41 & 60 \\
0                           & $1.00\times10^{-13}$ &  1.59 & 9.04  & 1.02 & 4.84 & 60 \\
0                           & $3.00\times10^{-13}$ &  1.71 & 9.53  & 0.94 & 6.22 & 60 \\
0                           & $4.00\times10^{-13}$ &  1.78 & 9.87  & 0.88 & 7.20 & 60 \\
0                           & $5.00\times10^{-13}$ &  1.90 & 10.43 & 0.80 & 8.85 & 60 \\
$1.00\times10^{-8}$         & $5.00\times10^{-13}$ &  1.90 & 10.43 & 0.80 & 8.85 & 60 \\
$1.00\times10^{-6}$         & $5.00\times10^{-13}$ &  1.90 & 10.43 & 0.80 & 8.85 & 60 \\ \hline

 \label{tab:t2}
\end{tabular}

\end{table}

\begin{table}

\caption{   Maximum gravitational
mass($M_{max}$) and the corresponding radius($R$) of  SQS  in dilaton gravity for various positive values of $\Lambda$ and$\alpha$ with $B_{bag}=75\frac{MeV}{fm^{3}}$.}

\begin{tabular}{lcccccr}
\hline
${\alpha}$  &  ${\Lambda}$  &  ${M_{max}(M_{\odot})}$ & ${R(Km)}$  &  ${\overline{\rho}}(10^{15}\frac{gr}{cm^{3}})$  &  ${z(10^{-1})}$  &  ${B_{bag}(\frac{MeV}{fm^{3}})}$ \\ \hline

$1.00\times10^{-7}$ & $1.00\times10^{-15}$ &  1.43 & 8.15 & 1.25 & 4.40 & 75 \\
$1.00\times10^{-7}$ & $1.00\times10^{-14}$ &  1.44 & 8.15 & 1.26 & 4.48 & 75 \\
$1.00\times10^{-8}$ & $1.00\times10^{-13}$ &  1.46 & 8.30 & 1.21 & 4.77 & 75 \\
0                   & $1.00\times10^{-13}$ &  1.46 & 8.30 & 1.21 & 4.77 & 75 \\
0                   & $3.00\times10^{-13}$ &  1.55 & 8.63 & 1.15 & 5.90 & 75 \\
0                   & $4.00\times10^{-13}$ &  1.60 & 8.87 & 1.09 & 6.60 & 75 \\
0                   & $5.00\times10^{-13}$ &  1.67 & 9.18 & 1.03 & 7.60 & 75 \\
$1.00\times10^{-8}$ & $5.00\times10^{-13}$ &  1.67 & 9.18 & 1.03 & 7.60 & 75 \\
$1.00\times10^{-6}$ & $5.00\times10^{-13}$ &  1.67 & 9.18 & 1.03 & 7.60 & 75 \\ \hline

\\ \label{tab:t3}
\end{tabular}

\end{table}


\begin{table}

   \caption{The percentage increase
of maximum mass of SQS  in dilaton gravity for various values of $\Lambda$ at $b=10^{-4}$ and $\alpha=0$ with different values of $B_{bag}$. }
\begin{tabular}{lcccr}
\hline
$\mathbf{B_{bag}}$&$\mathbf{{(\Delta M)}_{\Lambda_{1}}}$${[M_{max\Lambda_{1}}]}$&$\mathbf{{(\Delta M)}_{\Lambda_{2}}}$${[M_{max\Lambda_{2}}]}$&$\mathbf{{(\Delta M)}_{\Lambda_{3}}}$${[M_{max\Lambda_{3}}]}$&$\mathbf{{(\Delta M)}_{\Lambda_{4}}}$${[M_{max\Lambda_{4}}]}$ \\

&{ $(\Lambda_{1}=1.00\times10^{-13})$}&
{$(\Lambda_{2}=3.00\times10^{-13})$}&{$(\Lambda_{3}=4.00\times10^{-13})$}&{$(\Lambda_{4}=5.00\times10^{-13})$} \\ \hline
$90$&1.49\%&6.72\%&9.70\%&13.43\% \\
$75$&2.10\%&8.39\%&11.89\%&16.78\% \\
$60$&2.58\%&10.32\%&14.84\%&22.58\% \\ \hline
\end{tabular}
 \label{tab:t4}

\end{table}

\begin{table}

   \caption{   Maximum gravitational
mass($M_{max}$) and the corresponding radius($R$) of  SQS  in dilaton gravity for various negative values of $\Lambda$ at $b=10^{-4}$ and $\alpha=0$. }

\begin{tabular}{lccr}
\hline
${\alpha}$ & ${\Lambda}$ & ${M_{max}(M_{\odot})}$ & ${R(Km)}$
\\ \hline
0 &                     0 & 1.34 & 7.57 \\
0 & $-8.00\times10^{-13}$ & 1.19 & 6.94 \\
0 & $-1.00\times10^{-12}$ & 1.16 & 6.83 \\ \hline

\\
 \label{tab:t5}
\end{tabular}

\end{table}
\newpage


\clearpage



\clearpage

\clearpage

\clearpage

\end{document}